\documentclass[preprintnumbers,amsmath,amssymb,showpacs]{revtex4}
\usepackage{epsfig}
\usepackage{color}
\begin{document}
\setlength{\voffset}{1.0cm}
\title{Evidence for factorized scattering of composite states in the Gross-Neveu model}
\author{Christian Fitzner\footnote{fitzner@theorie3.physik.uni-erlangen.de}}
\author{Michael Thies\footnote{thies@theorie3.physik.uni-erlangen.de}}
\affiliation{Institut f\"ur Theoretische Physik III,
Universit\"at Erlangen-N\"urnberg, D-91058 Erlangen, Germany}
\date{\today}
\begin{abstract}
Scattering of two baryons in the large-$N$ Gross-Neveu model via the time-dependent Dirac-Hartree-Fock 
approach has recently been solved in closed analytical form. Here, we generalize this result to scattering processes
involving any number and complexity of the scatterers. The result is extrapolated from the solution of few
baryon problems, found via a joint ansatz for the scalar mean field and the Dirac spinors, and presented in 
analytical form. It has been verified numerically for up to 8-baryon problems so far, but a full mathematical proof is
still missing. Examples shown include the analogue of proton-nucleus and  nucleus-nucleus scattering in
this toy model. All the parameters of the general result can be fixed by one- and two-baryon input only. We take this finding
as evidence for factorized scattering, but on the level of composite multi-fermion states rather than elementary fermions.
\end{abstract}
\maketitle
\section{Introduction}\label{sect1}
The massless Gross-Neveu (GN) model \cite{L1} is the 1+1 dimensional quantum field theory of $N$ flavors of massless Dirac fermions, 
interacting through a scalar-scalar contact interaction. Suppressing flavor labels as usual, its Lagrangian reads
\begin{equation}
{\cal L} =   \bar{\psi} i\partial \!\!\!/ \psi + \frac{g^2}{2}(\bar{\psi} \psi)^2.
\label{1.1}
\end{equation}
The physics phenomena inherent in this simple looking Lagrangian are particularly rich and accessible in the 't~Hooft limit
($N \to \infty,\  Ng^2=$ const.), to which we restrict ourselves from here on. The GN model can be thought of as relativistic version
of particles moving along a line and interacting via an attractive $\delta$-potential. However, it exhibits many non-trivial
features characteristic for relativistic quantum fields such as covariance, renormalizability, asymptotic freedom, dimensional transmutation,
spontaneous symmetry breaking, interacting Dirac sea. It is also one of the few models known where most of the non-perturbative questions
of interest to strong interaction physics can be answered in closed analytical form. Such calculations have turned out to be both challenging 
and instructive, generating a continued interest in this particular ``toy model" over several decades, see e.g. the review articles \cite{L2,L3,L4}.

In the present paper we address the problem of time-dependent scattering of multi-fermion bound states in full generality. As will be recalled
in more detail in the next section, the GN model possesses bound states which can be viewed as ``baryons", with fermions bound in a
dynamically created ``bag" of the scalar field $\bar{\psi}\psi$ \cite{L5}. There are even multi-baryon bound states which might
be identified with ``nuclei" \cite{L3}. Standard large $N$ arguments tell us that all of these bound states can be described adequately 
within a relativistic version of the Hartree-Fock (HF) approach.

Turning to the baryon-baryon scattering problem, the tool of choice 
is the time-dependent version of Hartree-Fock (TDHF), as originally suggested by Witten \cite{L6}. The basic equations in that case
are easy to state,
\begin{equation}
( i \partial \!\!\!/ - S )  \psi_{\alpha} = 0,  \qquad S = - g^2 \sum_{\beta}^{\rm occ}
\bar{\psi}_{\beta}\psi_{\beta},
\label{1.2}
\end{equation}
but hard to solve, even in 1+1 dimensions. One of the reasons is the fact that the sum over occupied states includes the Dirac sea, 
so that one is dealing with an infinite set of coupled, non-linear partial differential equations. 
No systematic, analytical method for solving such a complicated problem is known. 
Nevertheless, the exact solution for the time-dependent scattering problem of two baryons has recently been found in closed analytical form
by means of a joint ansatz for $S$ and $\psi_{\alpha}$ \cite{L7}.
It provides us with a microscopic solution of the scattering of two composite, relativistic objects, exact in the large $N$ limit. The
necessary details will be briefly summarized below. This result encourages us to go on
and try to solve more ambitious scattering problems involving any number of bound states, including ``nuclei" in addition to the
``nucleons"  considered so far. 

The paper is organized as follows. 
In Sec.~\ref{sect2} we briefly summarize what is known about multi-fermion bound states and their interactions in the GN model.
We also remind the reader how the baryon-baryon scattering problem has been solved recently, since we shall use the same 
strategy in the present work. Sec.~\ref{sect3} is devoted to the Dirac equation and the ansatz for scalar potential and continuum spinors.
Secs.~\ref{sect4} and \ref{sect5} contain the central results of this work, namely the coefficients entering the ansatz, presented in the form
of an algorithm. In Sec.~\ref{sect6}, we explain the extent to which the general result has been checked so far. Sec.~\ref{sect7} deals
with the bound state spinors which are then used in Sec.~\ref{sect8} to discuss the issue of self-consistency and the fermion density.
Sec.~\ref{sect9} addresses scattering observables like time delays or deformations of bound states. Sec.~\ref{sect10} contains a few illustrative 
examples, followed by a short summary and outlook in Sec.~\ref{sect11}. 

\section{State of the art}\label{sect2}
To put this study into perspective, we summarize what is known about multi-fermion bound states and their mutual
interactions in the massless GN model, Eq.~(\ref{1.1}).
\subsection{Static solutions}\label{sect2a}
\begin{figure}[h]
\begin{center}
\epsfig{file=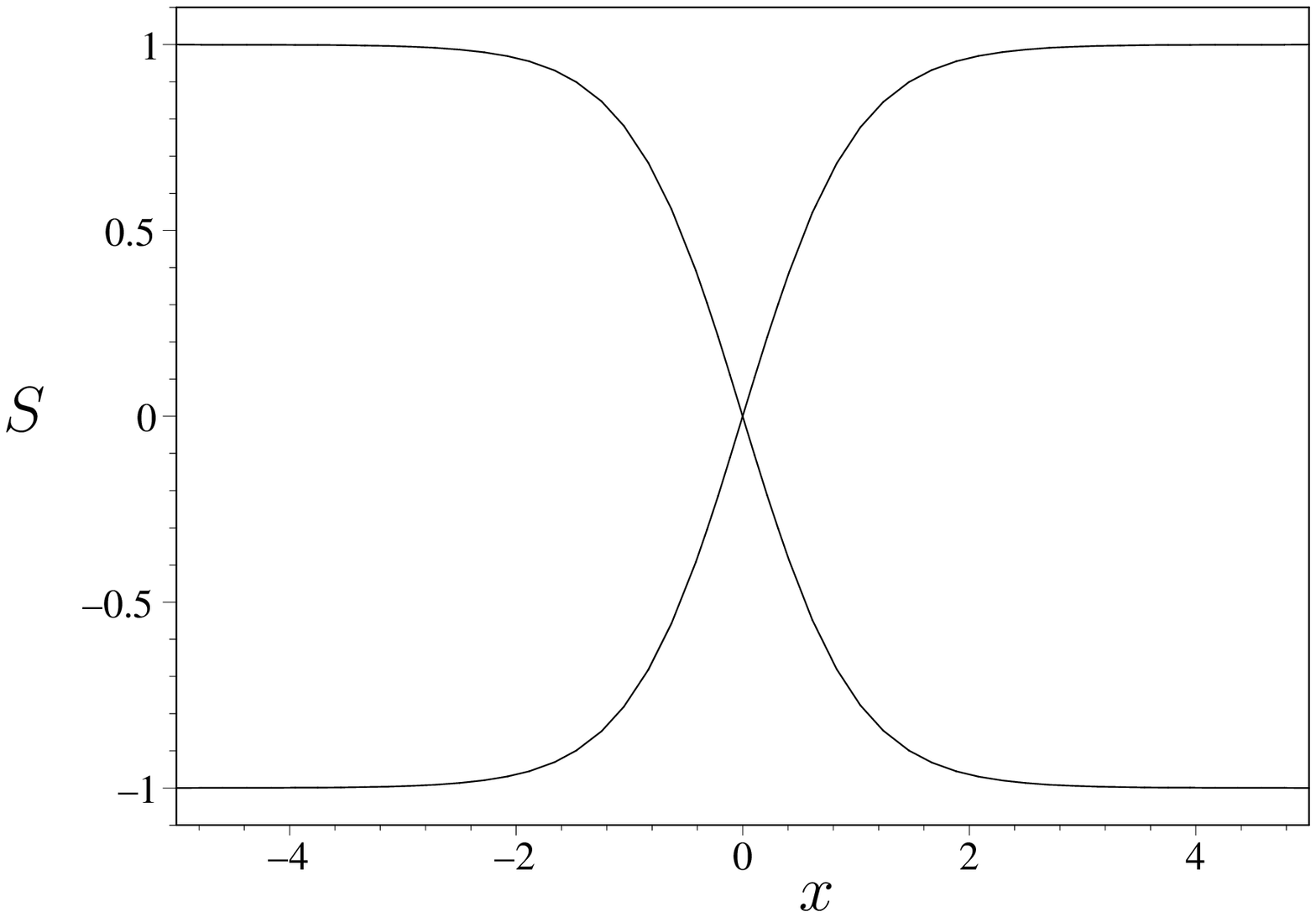,angle=270,width=8cm}
\caption{Scalar potential of kink (rising) and antikink (descending) in the GN model, interpolating between the two degenerate vacua $S=\pm 
1$ (units $m=1$).}
\label{fig1}
\end{center}
\end{figure}

\begin{figure}[h]
\begin{center}
\epsfig{file=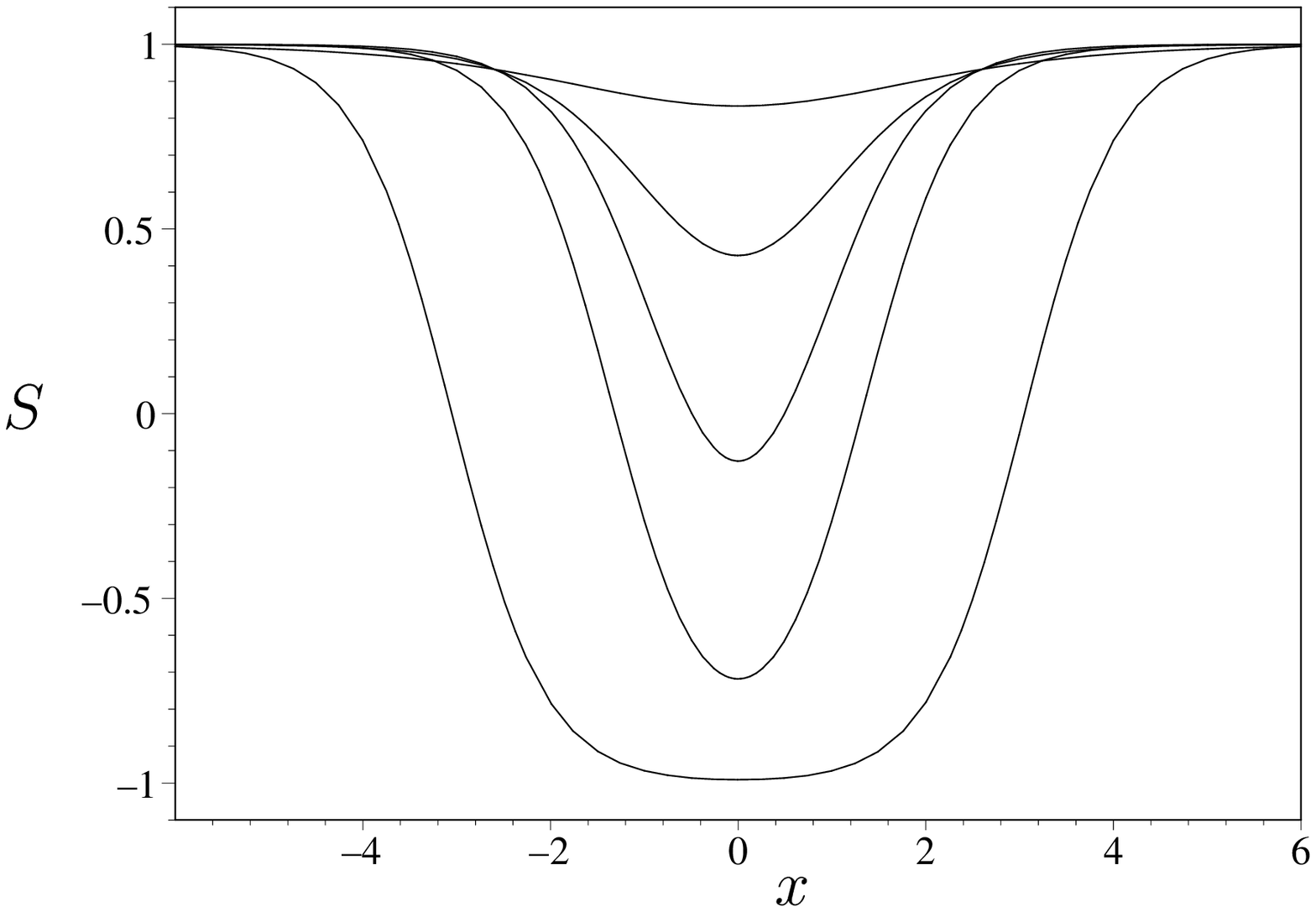,angle=270,width=8cm}
\caption{Scalar potential of DHN baryons in the GN model. Values of the parameter $y$ are 0.4, 0.7, 0.9, 0.99, 0.9999, from top to bottom.}
\label{fig2}
\end{center}
\end{figure}

Static multi-fermion bound states have been derived systematically with the help of inverse scattering theory and resolvent methods \cite{L3}.
The best known examples are the Callan-Coleman-Gross-Zee kink (cited in \cite{L8}) and the Dashen-Hasslacher-Neveu (DHN) baryon 
\cite{L5}, both of which can accommodate up to $N$ fermions. The kink is topologically non-trivial, reflecting the $Z_2$ chiral symmetry of 
the massless GN model. Its shape (shown in Fig.~\ref{fig1}) and mass are independent of its fermion content. The DHN baryon is topologically 
trivial and stabilized by the bound fermions which affect its shape and mass,
as illustrated in Fig.~\ref{fig2}.

\begin{figure}[h]
\begin{center}
\epsfig{file=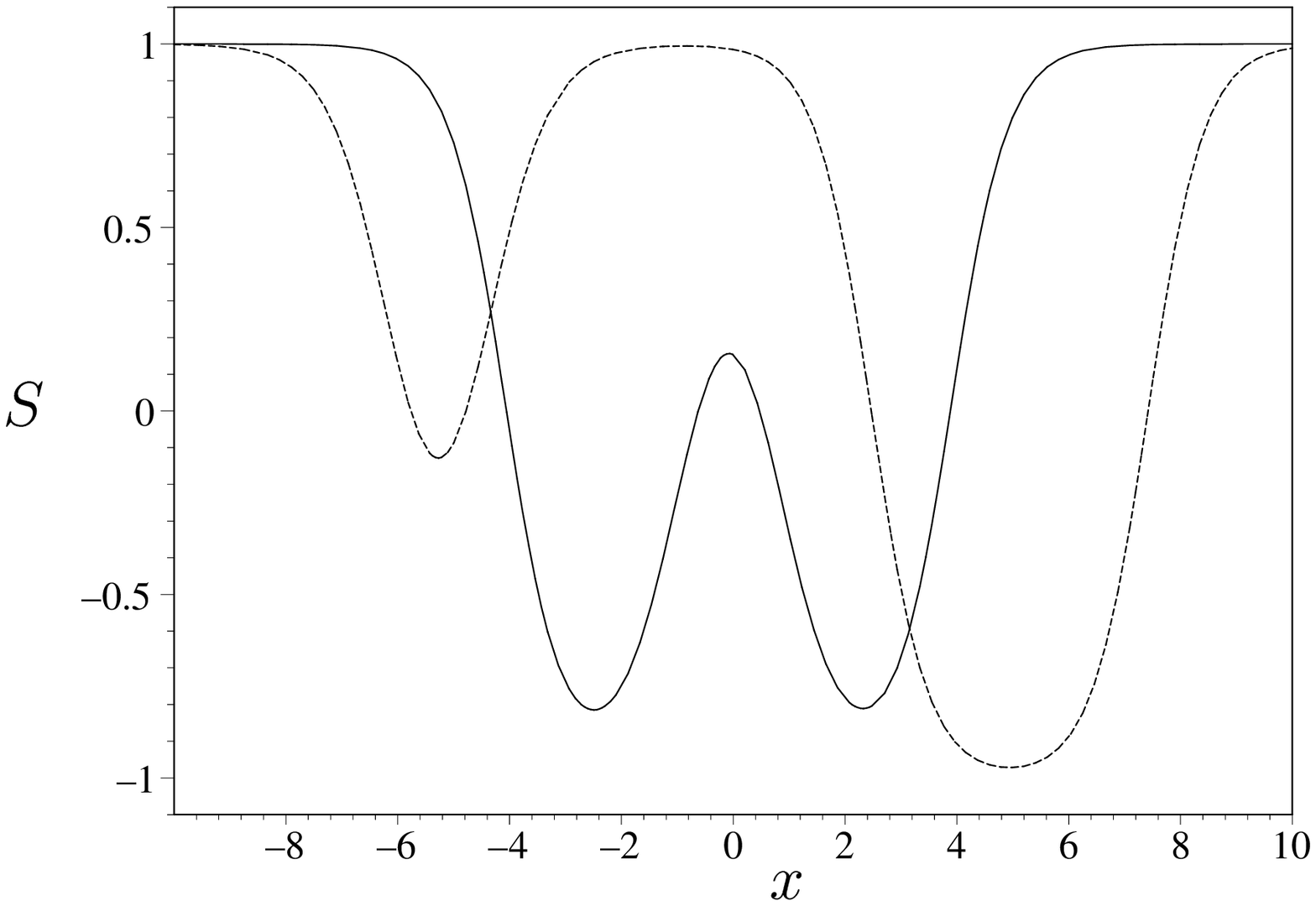,angle=270,width=8cm}
\caption{Examples of (topologically trivial) 2-baryon bound states in the GN model. $y$ parameters: 0.9999 and 0.9. The two curves differ in the 
relative position of the baryons ($\lambda_1=22.6,\lambda_2=0.06$ for the symmetric, $\lambda_1=0.018,\lambda_2=36.6$ for the asymmetric 
shape).}
\label{fig3}
\end{center}
\end{figure}

\begin{figure}[h]
\begin{center}
\epsfig{file=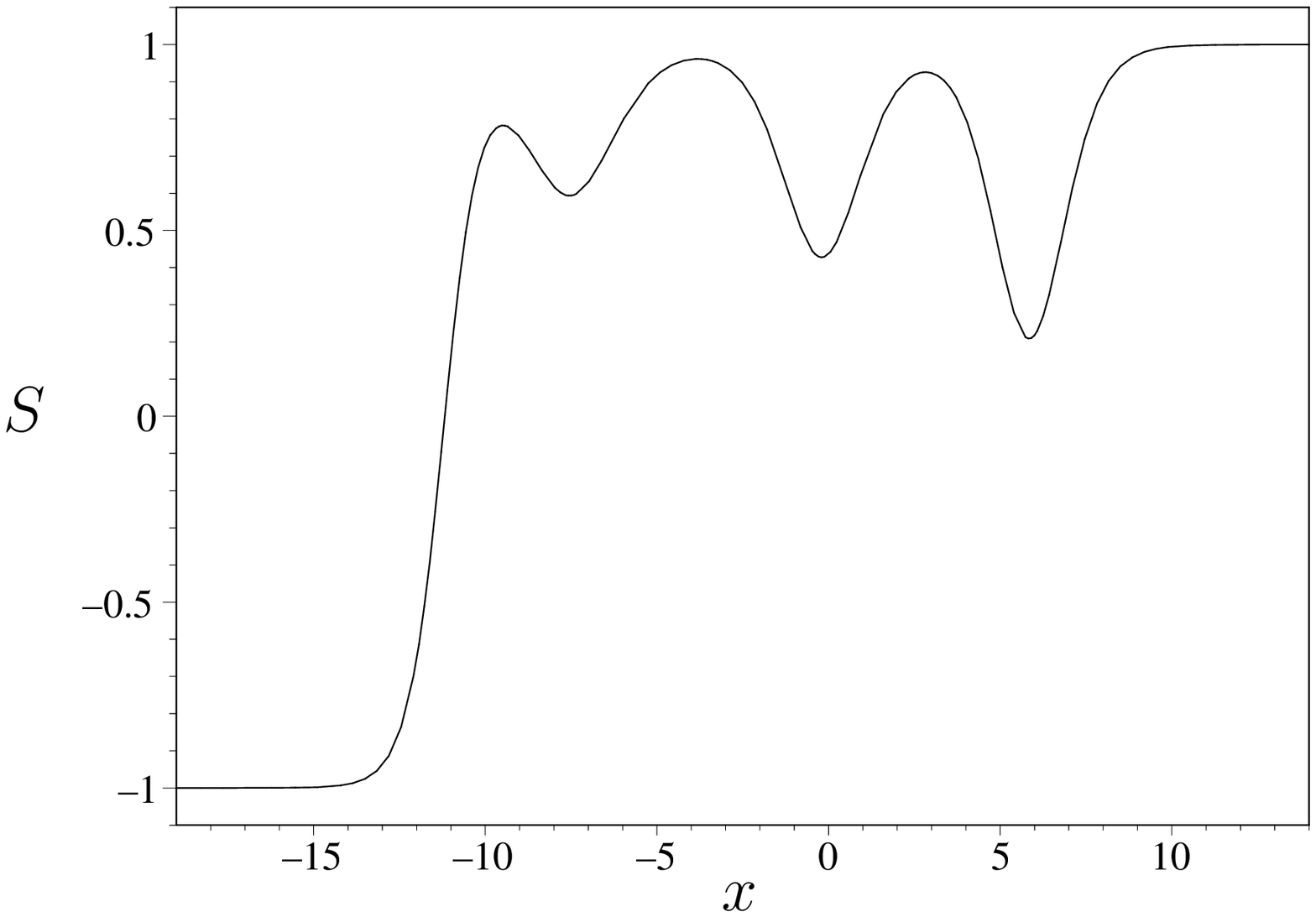,angle=270,width=8cm}
\caption{Example of a topologically non-trivial bound state of a kink and 3 DHN baryons. $y$ parameters: 1, 0.8, 0.7, 0.6.}
\label{fig4}
\end{center}
\end{figure}

Multi-baryon bound states have been constructed systematically by Feinberg \cite{L3}. They possess continuous 
parameters related to the position of the baryon constituents on which the mass of the bound state does not depend (``moduli"). 
They may be topologically trivial like the DHN baryon or non-trivial like the kink, depending on the (spatial) asymptotic behavior of $S$.
Some examples are shown in Figs.~\ref{fig3} and \ref{fig4}. 
A common feature of all static solutions is the fact that the scalar potential is transparent, i.e., the fermion reflection 
coefficient vanishes for all energies. Consequently the self-consistent, static solutions of the GN model coincide with the 
transparent scalar potentials of the Dirac equation, investigated independently by Nogami and coworkers \cite{L9,L10}. 
Since the static Dirac equation can be mapped onto a pair of (supersymmetric) Schr\"odinger 
equations, this also yields a bridge between static, self-consistent Dirac-HF solutions on the one hand and transparent potentials
of the Schr\"odinger equation on the other hand, a problem solved long ago by Kay and Moses \cite{L11}. The non-relativistic limit of the
topologically trivial, static GN solutions are well-known multi-soliton solutions of coupled non-linear Schr\"odinger (NLS) equations, arising in 
the Hartree approximation to particles in 1D with attractive $\delta$-interactions \cite{L12}. 

By boosting any static solution, one can trivially generate solutions of the TDHF equation \cite{L13}. This kind of solution enters 
in the asymptotic states of the scattering problem which we are going to study.
\subsection{Breather}\label{sect2b}
The breather is a time-dependent, oscillating solution of kink-antikink type. It was found by DHN, using the analogy with 
the sine-Gordon breather \cite{L5}. Since it is neither a conventional bound state nor a scattering state, it has no analogue in real particle
physics, but is reminiscent of collective, vibrational excitations of heavy nuclei or molecules. This underlines the classical character of the
large $N$ limit. We shall not consider scattering of breathers in the present work.
\subsection{Kink dynamics}\label{sect2c}
\begin{figure}[h]
\begin{center}
\epsfig{file=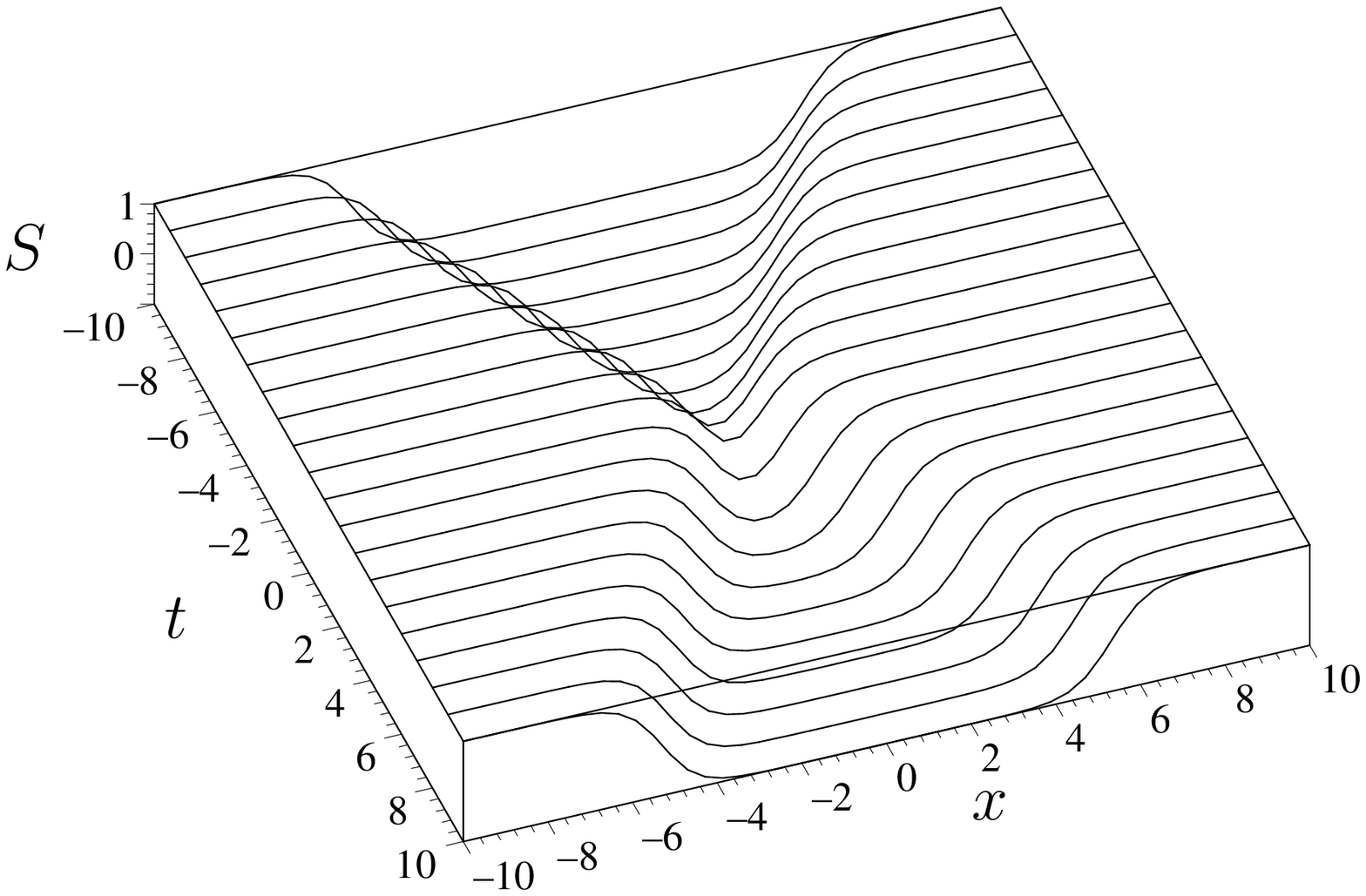,angle=270,width=8cm}
\caption{Time evolution of scalar potential for kink-antikink scattering at velocity $v=\pm 0.5$ \cite{L14}.}
\label{fig5}
\end{center}
\end{figure}

Following a suggestion in Ref.~\cite{L5}, kink-antikink scattering was solved in TDHF by analytic continuation of the breather \cite{L14}. 
Since the fermions do not react back, it is possible to map this problem rigorously onto the problem of kink-antikink
scattering in sinh-Gordon theory. If we set $S^2=e^{\theta}$, then $\theta$ satisfies the classical sinh-Gordon equation
\begin{equation}
\partial_{\mu}\partial^{\mu} \theta + 4 \sinh \theta = 0
\label{2.1}
\end{equation}
(in natural units), as first noticed by Neveu and Papanicolaou \cite{L15}. This mapping can be generalized. The known multi-soliton solutions
of the sinh-Gordon equation yield the self-consistent scalar potential for scattering of any number of kinks and antikinks \cite{L16}. 
A poor man's simulation of nuclear interactions was the scattering of ``trains" of solitons moving with almost the 
same speed in Ref.~\cite{L16} (there are no multi-soliton bound states). Kink dynamics has no non-relativistic analogue since the internal 
structure
of kink is ultrarelativistic, as evidenced by a  zero-energy bound state. Time-dependent kink-antikink scattering is illustrated in Fig.~\ref{fig5}. 

A crucial ingredient in proving the correspondence between kink dynamics and sinh-Gordon solitons is the fact that kink solutions satisfy 
the self-consistency mode-by-mode. They are of ``type I" in the classification of \cite{L14}, i.e., $\bar{\psi_{\alpha}}\psi_{\alpha} = 
\lambda_{\alpha}S$ with constant $\lambda_{\alpha}$ for every single particle state $\alpha$. This is also the basis for an interesting 
geometrical interpretation
of TDHF solutions,
relating time-dependent solutions of the GN model to the embedding of surfaces of constant mean curvature into 3D spaces \cite{L17}.

\subsection{Baryon-baryon scattering}\label{sect2d}
\begin{figure}[h]
\begin{center}
\epsfig{file=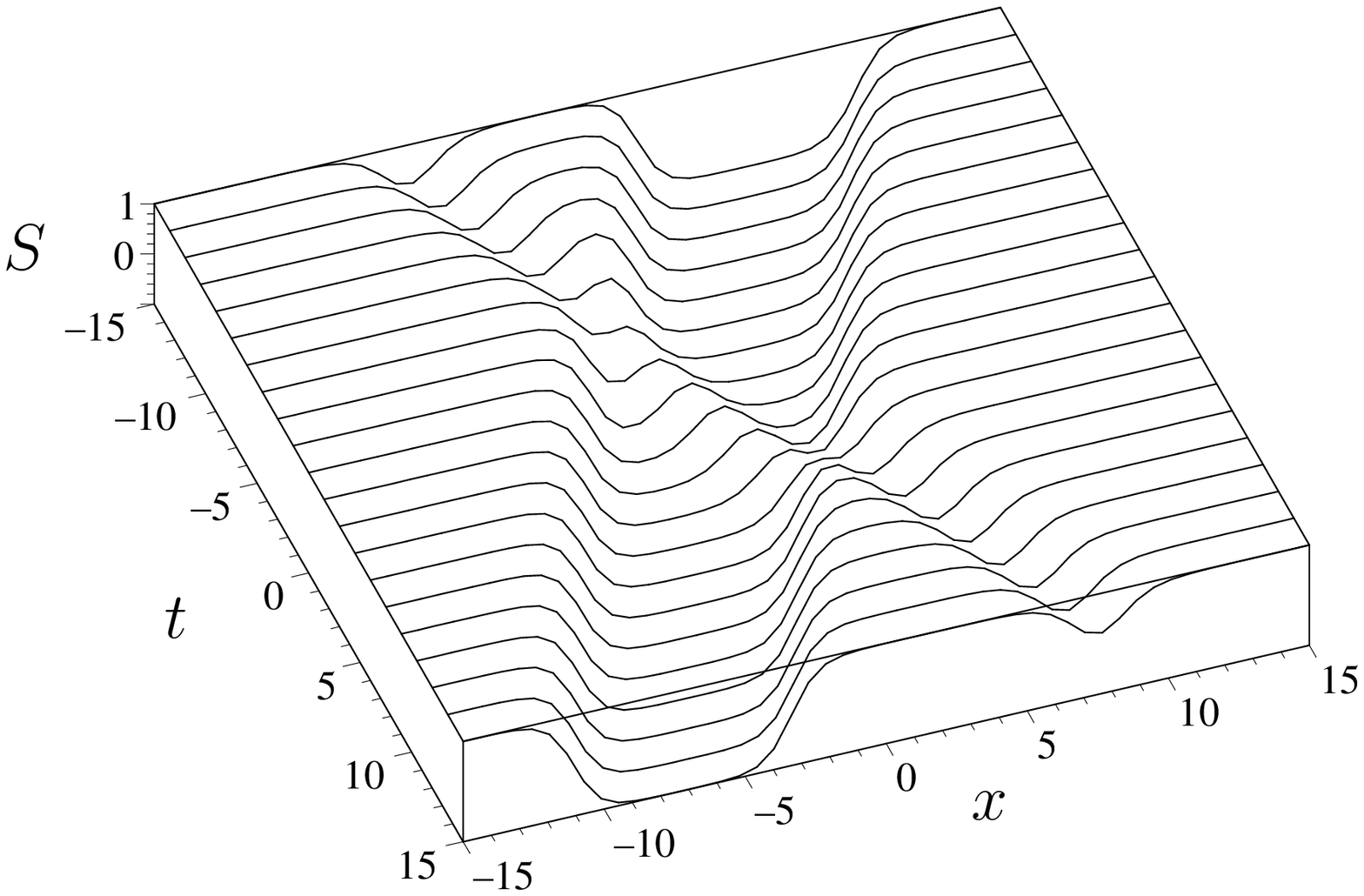,angle=270,width=8cm}
\caption{Time evolution of scalar potential for baryon-baryon scattering (parameters: $y_1=0.8, y_2=1-10^{-7}, v=\pm 0.4$) \cite{L7}.}
\label{fig6}
\end{center}
\end{figure}

Scattering of DHN baryons is significantly more involved than kink-antikink scattering. Presumably because the fermions react back, it does
not seem possible to map this problem onto any known soliton equation. The exact TDHF solution for baryon-baryon scattering was found 
recently in a different way, namely by ansatz \cite{L7}.  A specific example is illustrated in Fig.~\ref{fig6}. 
Since we shall follow the same strategy in the present paper, we briefly recall the main ideas behind the ansatz, referring the reader to 
Ref. \cite{L7} for technical details. 

The ansatz can best be described as follows. We start from the scalar mean field of a single 
(boosted) DHN baryon with label $i$. It can be cast into the form of a rational function of an exponential $U_i$,
\begin{equation}
S_i = \frac{1 + a_1^i U_i + U_i^2}{1+b_1^i U_i + U_i^2}, \qquad U_i = \lambda_i \exp\left\{ 2 y_i \gamma_i (x-v_i t) \right\}.
\label{2.2}
\end{equation} 
Here, $y_i$ is a parameter governing the size of the baryon and related to its fermion number $n_i$ via 
\begin{equation}
y_i = \sin \frac{\pi n_i}{2 N},
\label{2.3}
\end{equation}
$v_i$ denotes the baryon velocity, $\gamma_i = (1-v_i^2)^{-1/2}$, and $\lambda_i$ is an arbitrary real factor expressing the freedom of 
choosing the initial baryon position.
The Dirac components of the continuum spinor have the same rational form with different coefficients in the numerator only and an 
additional 
plane wave factor,
\begin{equation}
\psi_k  =  \left( \begin{array}{c} c_0^i + c_1^i U_i + c_2^i U_i^2 \\ d_0^i + d_1^i U_i + d_2^i U_i^2 \end{array} \right) \frac{e^{i(kx-\omega t)}}
{1+b_1^i U_i + U_i^2}.
\label{2.4}
\end{equation}
The asymptotic behavior at fixed $t$ is $\psi_k \sim e^{ikx}$ for $x\to \pm \infty$, showing that the potential is transparent.  

In order to solve the scattering problem for baryons $i$ and $j$, we start by multiplying $S_i$ and $S_j$ and expand the numerator and 
denominator,
\begin{equation}
S_i S_j  =   \frac{1 + a_1^i U_i + a_1^j U_j + U_i^2 + a_1^i a_1^j U_i U_j + U_j^2 + a_1^j U_i^2 U_j + a_1^i U_i U_j^2 + U_i^2 U_j^2}
{1 + b_1^i U_i + b_1^j U_j + U_i^2 + b_1^i b_1^j U_i U_j + U_j^2 + b_1^j U_i^2 U_j + b_1^i U_i U_j^2 + U_i^2 U_j^2}.
\label{2.5}
\end{equation}
This may be viewed as scalar potential for non-interacting baryons. 
The ansatz for interacting baryons proposed in \cite{L7} now consists in assuming
that the only effect of the interaction is to change the coefficients in the numerator and denominator of (\ref{2.5}), keeping the polynomial
dependence on $U_i,U_j$
the same. Likewise, the ansatz for the spinor is obtained by multiplying the rational factors of $\psi_k$ for baryons $i$ and $j$ and allowing
for changes in the coefficients
only. The overall exponential factor is kept unchanged, since it is expected that the potential is reflectionless also in the interacting case. 
It turns out that most of the coefficients in $S$ and $\psi_k$ are in fact determined by the asymptotic in- and out-states. Only 4 coefficients 
remain to be determined,
namely the factors in front of the monomials $U_iU_j$ in the three numerators and the common denominator.
Inserting this ansatz into the Dirac equation determines the missing coefficients and confirms that this simple idea yields the exact solution 
of the 2-baryon problem. 

So far, we have discussed only the fermion continuum states. Bound states can be obtained by analytic continuation in a spectral parameter
(a function of $k,\omega$) and subsequent normalization. Self-consistency can then be checked explicitly, confirming that the ansatz solves
the TDHF problem. 
The solution is found to be of type III, i.e., the scalar density of any single particle orbit can be expressed as a linear combination of 3 distinct
functions of 
($x,t$).
We have no a priori argument why the ansatz should be successful, but its simple form is most certainly a large-$N$ manifestation of the 
quantum integrability of finite-$N$ GN models.

The result for the non-trivial coefficients is rather complicated, but by a proper choice of variables and light cone coordinates, one manages 
to keep all coefficients in rational form. Unlike in the kink-antikink case, the non-relativistic limit is now accessible, since the DHN baryon goes
over into the soliton of the NLS equation in the limit of small fermion number. Starting from the two-baryon solution, one then recovers the 
time-dependent solutions of the multi-component NLS equation of Nogami and Warke for $N=2$ \cite{L12}. 

This completes the overview of the present state of the art. Here we propose to extend the two-baryon TDHF scattering solution of 
Ref.~\cite{L7} 
to an arbitrary
number of composite colliding particles, including multi-baryon bound states (``nuclei") in addition to baryons. The central idea is to use an
ansatz for the scalar potential inspired by the product of $N$ single baryon potentials, assuming that only the coefficients of the resulting 
rational function of $U_1,...,U_N$ will be affected by the interactions.

\section{Ansatz and Dirac equation}\label{sect3}
A convenient choice of the Dirac matrices in 1+1 dimensions is
\begin{equation}
\gamma^0 = \sigma_1,\quad \gamma^1 = i \sigma_2, \quad \gamma_5 = \gamma^0 \gamma^1 = - \sigma_3.
\label{3.1}
\end{equation}
Together with light cone coordinates
\begin{equation}
z= x-t, \quad \bar{z}=x+t, \quad \partial_0 = \bar{\partial}-\partial,\quad \partial_1 = \bar{\partial} + \partial,
\label{3.2}
\end{equation}
this simplifies the Dirac-TDHF equation to
\begin{equation}
2i \bar{\partial} \psi_2 = S \psi_1, \quad 2i \partial \psi_1 = - S \psi_2 .
\label{3.3}
\end{equation}
Here, $\psi_1$ is the upper, left-handed, $\psi_2$ the lower, right-handed spinor component.
We posit the following ansatz for the scalar TDHF potential,
\begin{equation}
S= \frac{\cal N}{\cal D}.
\label{3.4}
\end{equation}
As motivated in the preceding section, $S$ is assumed to be a rational function of $N$ exponentials $U_i$, where $N$ is the number of baryons,
\begin{eqnarray}
{\cal N} & = & \sum_{\{i_k\}} a_{i_1...i_N}^{1...N} U_1^{i_1}...U_N^{i_N},
\nonumber \\
{\cal D} & = &  \sum_{\{i_k\}} b_{i_1...i_N}^{1...N} U_1^{i_1}...U_N^{i_N}.
\label{3.5}
\end{eqnarray}
Each summation index $i_k$ runs over the values 0,1,2, and the coefficients $a,b$ are real. 
The basic exponential $U_i$ has the form inferred from the single DHN baryon in flight, 
\begin{equation}
U_i = \lambda_i \exp \left\{ y_i \left( \eta_i^{-1} \bar{z} + \eta_i z \right) \right\}.
\label{3.6}
\end{equation}
The parameter $y_i$ specifies the size (or, equivalently, fermion number) of the $i$-th baryon. $\eta_i$ is related to the baryon rapidity 
$\xi_i$ and velocity $v_i$ via
\begin{equation}
\eta_i = e^{\xi_i} = \sqrt{\frac{1+v_i}{1-v_i}}.
\label{3.7}
\end{equation}
For $y_i$ we shall use the parametrization
\begin{equation}
y_i = \frac{Z_i^2-1}{2iZ_i}, \quad Z_i = i y_i - \sqrt{1-y_i^2}, \quad |Z_i|^2 = 1,
\label{3.8}
\end{equation}
to avoid the appearance of square roots. Apart from the $2N$ parameters $\{ Z_i, \eta_i\}$, the baryon constituents are characterized by 
$N$ arbitrary, real scale factors $\lambda_i$ needed to specify their initial positions. The $U_i$ must be ordered according to baryon velocities.
We choose the convention that $v_i \ge v_j$ if $i<j$.

We now turn to the ansatz for the continuum spinors, assuming from the outset that the TDHF potential is reflectionless, 
\begin{equation}
\psi_{\zeta} = \left( \begin{array}{c} \zeta {\cal N}_1 \\ -{\cal N}_2 \end{array} \right) \frac{e^{i(\zeta \bar{z}-z/\zeta)/2}}{{\cal D} \sqrt{\zeta^2+1}}.
\label{3.9}
\end{equation}
Here, $\zeta$ denotes the light cone spectral parameter related to ordinary momentum and energy via
\begin{equation}
k= \frac{1}{2} \left( \zeta- \zeta^{-1} \right), \quad \omega =  - \frac{1}{2} \left( \zeta + \zeta^{-1} \right).
\label{3.10}
\end{equation}
${\cal N}_1,{\cal N}_2$ are multivariate polynomials in the $U_i$ of the same degree as ${\cal N},{\cal D}$, 
\begin{eqnarray}
{\cal N}_1 & = & \sum_{\{i_k\}} c_{i_1...i_N}^{1...N} U_1^{i_1}...U_N^{i_N},
\nonumber \\
{\cal N}_2 & = &  \sum_{\{i_k\}} d_{i_1...i_N}^{1...N} U_1^{i_1}...U_N^{i_N},
\label{3.11}
\end{eqnarray}
but now with complex coefficients $c,d$. In Eq.~(\ref{3.9}) we have factored out the free Dirac spinor
\begin{equation}
\psi_{\zeta}^{(0)} =  \left( \begin{array}{c} \zeta  \\ - 1  \end{array} \right) \frac{e^{i(\zeta \bar{z}-z/\zeta)/2}}{\sqrt{\zeta^2+1}}
\label{3.12}
\end{equation}
to ensure that all polynomials start with a ``1". The denominator ${\cal D}$ in the spinor, Eq.~(\ref{3.9}), is assumed to be the same as in the 
scalar potential, Eq.~(\ref{3.4}).
Inserting this ansatz into the Dirac equation (\ref{3.3}) yields 
\begin{eqnarray}
0 & = & 2i \zeta^{-1} \left({\cal N}_2\bar{\partial} {\cal D} - {\cal D} \bar{\partial} {\cal N}_2 \right) +  {\cal N}_2 {\cal D} - {\cal N}_1{\cal N} ,
\nonumber \\
0 & = & 2i \zeta \left({\cal D}\partial {\cal N}_1 - {\cal N}_1 \partial {\cal D} \right) +  {\cal N}_1 {\cal D} - {\cal N}_2{\cal N}.
\label{3.13}
\end{eqnarray}
Actually, we can eliminate the variable $\zeta$ by rescaling $z,\bar{z}$ via $z \to \zeta z, \bar{z} \to \zeta^{-1}\bar{z}$. This transforms $U_i$ into
\begin{equation}
U_i = \lambda_i \exp \left\{ y_i \left( \zeta_i^{-1} \bar{z} + \zeta_i z \right) \right\}, \quad \zeta_i = \eta_i \zeta.
\label{3.14}
\end{equation} 
The final form of the Dirac equation can then be obtained  by setting $\zeta=1$ in Eq.~(\ref{3.13}),
\begin{eqnarray}
0 & = & 2i  \left({\cal N}_2\bar{\partial} {\cal D} - {\cal D} \bar{\partial} {\cal N}_2 \right) +  {\cal N}_2 {\cal D} - {\cal N}_1{\cal N} ,
\nonumber \\
0 & = & 2i  \left({\cal D}\partial {\cal N}_1 - {\cal N}_1 \partial {\cal D} \right) +  {\cal N}_1 {\cal D} - {\cal N}_2{\cal N}.
\label{3.15}
\end{eqnarray}

The numerator and denominator functions (${\cal N}, {\cal D}, {\cal N}_1, {\cal N}_2$) are polynomials in the $U_i$. Since the $U_i$ 
are eigenfunctions of $\partial, \bar{\partial}$, the Dirac equation (\ref{3.15}) gets converted into the condition that 2 
polynomials vanish identically. 
Thus each coefficient of the monomials $U_1^{i_1}...U_N^{i_N}$ must vanish separately. The number of terms in each of the polynomials, 
Eqs.~(\ref{3.5}) and (\ref{3.11}),
is $3^N$ for $N$ baryons, as $U_i$ can appear with powers 0,1,2. In the final Dirac equation, $U_i$ appears
with powers $0...4$, so that Eq.~(\ref{3.15}) is altogether equivalent to $2 \times 5^N$ algebraic equations
for the coefficients $a,b,c,d$ of our ansatz. 
\section{Reduction formulas and reducible coefficients}\label{sect4}
In this and the following section, we present our results for the coefficients entering the scalar potential and the continuum spinors for 
$N$ baryons, i.e., the coefficients
of the polynomials ${\cal N}, {\cal D}, {\cal N}_1, {\cal N}_2$ introduced above. They fall naturally into 2 classes: ``Reducible" coefficients
which can be related to the $N-1$ baryon 
problem, and ``irreducible" ones which cannot. The reducible coefficients are the subject of this section, the irreducible ones will be 
discussed in the next section.

There are two distinct ways of reducing the $N$ baryon problem to the $N-1$ baryon problem, either by letting $U_k\to 0$ or by letting 
$U_k \to \infty$.

In both cases, $U_k$ drops out of the expressions for $S$ and $\psi_{\zeta}$.
Since this can be done for any label $k$, one gets a large number of recursion relations. As explained in greater detail in Ref.~\cite{L7},
one has to take into account time delays and (in the case of the spinors) transmission amplitudes for final states, depending on whether the
eliminated baryon $k$ has been scattered from the remaining $N-1$ baryons or not.

Let us consider the scalar potential first. Starting point are the following basic relations,
\begin{eqnarray}
\lim_{U_k \to 0} S(U_1,...,U_N) & = & S(U_1,...,U_{k-1}, \delta_{k,k+1}U_{k+1},...,\delta_{kN}U_N),
\nonumber \\
\lim_{U_k \to \infty} S(U_1,...,U_N) & = & S(\delta_{k1}U_1,...,\delta_{k,k-1}U_{k-1},U_{k+1},...,U_N).
\label{4.1}
\end{eqnarray} 
$U_k$ is missing on the right hand side, which therefore refers to $N-1$ baryons. The $\delta_{ij}$ are (real) time delay factors satisfying 
\cite{L7} 
\begin{equation}
\delta_{ij}  = \frac{1}{\delta_{ji}} =  \frac{(\zeta_j Z_i+\zeta_i Z_j)(\zeta_i Z_i+\zeta_j Z_j) (\zeta_i Z_i Z_j-\zeta_j)(\zeta_j Z_i Z_j-\zeta_i)}
{(\zeta_j Z_i-\zeta_i Z_j)(\zeta_i Z_i-\zeta_j Z_j) (\zeta_i Z_i Z_j+\zeta_j)(\zeta_j Z_i Z_j+\zeta_i)} \qquad (i<j).
\label{4.2}
\end{equation}
It is important to keep track of the ordering of the baryon labels ($v_i \ge v_j$ if $i<j$) when applying these formulas.
Relations (\ref{4.1}) imply the following recursion relations for the coefficients in (\ref{3.5}),
\begin{eqnarray}
\left. a_{i_1...i_N}^{1...N}\right|_{i_k=0} & = & C_k a_{i_1...\overline{i_k}...i_N}^{1...\overline{k}...N} \prod_{\ell=k+1}^{N} \delta_{k\ell}^{i_{\ell}},	 
\nonumber \\
\left. b_{i_1...i_N}^{1...N} \right|_{i_k=0}  & = &  C_k b_{i_1...\overline{i_k}...i_N}^{1...\overline{k}...N} \prod_{\ell=k+1}^{N} \delta_{k\ell}^{i_{\ell}},
\nonumber \\
\left. a_{i_1...i_N}^{1...N} \right|_{i_k=2} & = &  C_k' a_{i_1...\overline{i_k}...i_N}^{1...\overline{k}...N} \prod_{\ell=1}^{k-1} \delta_{k\ell}^{i_{\ell}},	
\nonumber \\
\left. b_{i_1...i_N}^{1...N} \right|_{i_k=2} & = &  C_k' b_{i_1...\overline{i_k}...i_N}^{1...\overline{k}...N} \prod_{\ell=1}^{k-1} \delta_{k\ell}^{i_{\ell}}.
\label{4.3}
\end{eqnarray}
We use the convention that barred indices have to be omitted. The factors $C_k, C_k'$ appear here because relations (\ref{4.1}) 
determine only the 
ratio ${\cal N}/{\cal D}$. They can be fixed as follows. We normalize the lowest and highest coefficients of ${\cal N},{\cal D}$ to
1 for any number of baryons,
\begin{eqnarray}
a_{0...0}^{1...N} & = & 1, \quad b_{0...0}^{1...N} \ = \ 1,
\nonumber \\
a_{2...2}^{1...N} & = & 1, \quad b_{2...2}^{1...N} \ = \ 1.
\label{4.4}
\end{eqnarray} 
This is always possible since we must recover the vacuum potential $S=1$ in the limit where all $U_i$ go to 0 or $\infty$, and the $U_i$ 
contain arbitrary
scale factors $\lambda_i$, see Eq.~(\ref{3.6}). Specializing relations (\ref{4.3}) to the cases where all indices are 0 or all indices are 2 and 
using Eq.~(\ref{4.2}), we then find
\begin{eqnarray}
C_k & = &  1,
\nonumber \\
C_k' & = & \prod_{\ell =1}^{k-1} \delta_{\ell k}^2 .
\label{4.5}
\end{eqnarray}
This yields the following final recursion relations for the coefficients entering $S$, 
\begin{eqnarray}
\left. a_{i_1...i_N}^{1...N} \right|_{i_k=0} & = &  a_{i_1...\overline{i_k}...i_N}^{1...\overline{k}...N} \prod_{\ell=k+1}^{N} \delta_{k\ell}^{i_{\ell}},	 
\nonumber \\
\left. b_{i_1...i_N}^{1...N} \right|_{i_k=0} & = &  b_{i_1...\overline{i_k}...i_N}^{1...\overline{k}...N} \prod_{\ell=k+1}^{N} \delta_{k\ell}^{i_{\ell}},
\nonumber \\
\left. a_{i_1...i_N}^{1...N} \right|_{i_k=2} & = &  a_{i_1...\overline{i_k}...i_N}^{1...\overline{k}...N} \prod_{\ell=1}^{k-1} \delta_{\ell k}^{2-i_{\ell}},	
\nonumber \\
\left. b_{i_1...i_N}^{1...N} \right|_{i_k=2} & = &  b_{i_1...\overline{i_k}...i_N}^{1...\overline{k}...N} \prod_{\ell=1}^{k-1} \delta_{\ell k}^{2-i_{\ell}}.
\label{4.6}
\end{eqnarray}
They determine all $N$-baryon coefficients containing at least one 0 or one 2 in their subscripts in terms of $(N-1)$-baryon coefficients,
leaving only the two irreducible coefficients
$a_{11...1}^{12...N}, b_{11...1}^{12...N}$ in front of $U_1...U_N$ undetermined. 

For the spinors, we have to take into account 
transmission amplitudes in addition to the time delay factors. Consequently the general reduction formulas (\ref{4.1}) have to
be replaced by
\begin{eqnarray}
\lim_{U_k \to 0}\psi_{\zeta}(U_1,...,U_N) & = & \psi_{\zeta}(U_1,...,U_{k-1}, \delta_{k,k+1}U_{k+1},...,\delta_{k N}U_N),
\nonumber \\
\lim_{U_k \to \infty} \psi_{\zeta}(U_1,...,U_N) & = & T_k \psi_{\zeta}(\delta_{k 1}U_1,...,\delta_{k,k-1}U_{k-1},U_{k+1},...,U_N),
\label{4.7}
\end{eqnarray} 
where $T_k$ is the transmission amplitude of baryon $k$ \cite{L7}
\begin{equation}
T_k  =  \frac{(\zeta_k+Z_k)(\zeta_kZ_k-1)}{(\zeta_k-Z_k)(\zeta_kZ_k+1)}.
\label{4.7a}
\end{equation}
It is unitary ($|T_k|=1$) due to the reflectionless potential. Using a normalization analogous to (\ref{4.4}), i.e.,
\begin{eqnarray}
c_{0...0}^{1...N} & = & 1, \quad d_{0...0}^{1...N} \ = \ 1,
\nonumber \\
c_{2...2}^{1...N} & = & T_1...T_N, \quad d_{2...2}^{1...N} \ = \ T_1...T_N,
\label{4.8}
\end{eqnarray} 
we arrive at the recursion relations
\begin{eqnarray}
\left. c_{i_1...i_N}^{1...N}\right|_{i_k=0} & = &  c_{i_1...\overline{i_k}...i_N}^{1...\overline{k}...N} \prod_{\ell=k+1}^{N} \delta_{k \ell}^{i_{\ell}},	 
\nonumber \\
\left. d_{i_1...i_N}^{1...N}\right|_{i_k=0} & = &  d_{i_1...\overline{i_k}...i_N}^{1...\overline{k}...N} \prod_{\ell=k+1}^{N} \delta_{k \ell}^{i_{\ell}},
\nonumber \\
\left. c_{i_1...i_N}^{1...N}\right|_{i_k=2} & = &  c_{i_1...\overline{i_k}...i_N}^{1...\overline{k}...N} T_k \prod_{\ell=1}^{k-1} \delta_{\ell k}^{2-i_{\ell}},	
\nonumber \\
\left. d_{i_1...i_N}^{1...N}\right|_{i_k=2} & = &  d_{i_1...\overline{i_k}...i_N}^{1...\overline{k}...N} T_k \prod_{\ell=1}^{k-1} \delta_{\ell k}^{2-i_{\ell}},
\label{4.9}
\end{eqnarray}
for the coefficients in ${\cal N}_1, {\cal N}_2$. Once again this leaves only the two irreducible coefficients $c_{11...1}^{12...N}, d_{11...1}^{12...N}$ 
of $U_1...U_N$ undetermined. Altogether, there are
$4\times 3^N$ coefficients in the ansatz for $S$ and $\psi_{\zeta}$ for $N$ baryons. All but the 4 irreducible ones are determined by 
normalization and recursion relations.

The first step towards solving the $N$ baryon problem is to eliminate all reducible coefficients, expressing the 4 polynomials in terms 
of irreducible coefficients,
time delay factors and transmission amplitudes only. The above recursion scheme enables us to do just this. The result can most
conveniently be cast into the form of an algorithm. We first formulate the algorithm and subsequently illustrate it with the explicit
results for $N=2,3$ and point out its advantages.
The algorithm will be stated separately for the 4 polynomials ${\cal D}, {\cal N}, {\cal N}_1, {\cal N}_2$.
\begin{enumerate}
\item Denominator ${\cal D}$ of $S$
\begin{enumerate}
\item Write down the product
\begin{equation}
{\cal D} = \prod_{i=1}^N \left( V_i + W_i\right)
\label{4.10}
\end{equation}
and expand it.
\item If a term contains between 2 and $N$ factors $V$, replace it by  
\begin{equation}
V_iV_j...  \to  \frac{b_{11...}^{ij...}}{b_1^ib_1^j...} V_iV_j...
\label{4.11}
\end{equation}
\item Substitute
\begin{equation}
W_i \to 1+ \left( \frac{V_i}{b_1^i} \right)^2
\label{4.12}
\end{equation}
and expand again.
\item If any term contains $(V_i V_j)^n$ ($i < j, n=1,2$), replace it by 
\begin{equation}
(V_i V_j)^n \to \frac{(V_i V_j)^n}{\delta_{ij}^n}.
\label{4.13}
\end{equation}
\item Set
\begin{equation}
V_k  =  b_1^k U_k \prod_{\ell=1}^{k-1} \delta_{\ell k}.
\label{4.14}
\end{equation}
\end{enumerate}
\item Numerator ${\cal N}$ of $S$
\vskip 0.5cm
The numerator ${\cal N}$ of $S$ can be obtained from the denominator ${\cal D}$ of $S$ by replacing all $b$-coefficients by 
$a$-coefficients,
\begin{equation}
b_1^i \to a_1^i, \qquad b_{11}^{ij} \to a_{11}^{ij}, \qquad ...
\label{4.15}
\end{equation}
\item Numerator ${\cal N}_1$ of $\psi_1$ 
\vskip 0.5cm
To get ${\cal N}_1$, start from ${\cal D}$ and perform the following steps:
\begin{enumerate}
\item Replace 
\begin{equation}
U_i^2 \to T_i U_i^2 ,
\label{4.16}
\end{equation}
where $T_i$ is the transmission amplitude of baryon $i$.
\item 
Replace all $b$-coefficients by $c$-coefficients,  
\begin{equation}
b_1^i \to c_1^i, \qquad b_{11}^{ij} \to c_{11}^{ij}, \qquad ...
\label{4.17}
\end{equation}
\end{enumerate}
\item  Numerator ${\cal N}_2$ of $\psi_2$ 
\vskip 0.5cm
To get ${\cal N}_2$, start from ${\cal N}_1$ and replace all $c$-coefficients by $d$-coefficients,
\begin{equation}
c_1^i \to d_1^i, \qquad c_{11}^{ij} \to d_{11}^{ij}, \qquad ...
\label{4.18}
\end{equation}
\end{enumerate}
To avoid misunderstandings, we illustrate the outcome of the algorithm with a few explicit examples. For $N=2$ (9 terms), one finds
\begin{eqnarray}
{\cal D}  & = &  1 + b_1^1 U_1 + b_1^2 \delta_{12} U_2 + U_1^2 + b_{11}^{12} U_1 U_2 + \delta_{12}^2 U_2^2
 + b_1^2 U_1^2 U_2 + b_1^1 \delta_{12} U_1 U_2^2 + U_1^2 U_2^2 ,
\nonumber \\
{\cal N}  & = &  1 + a_1^1 U_1 + a_1^2 \delta_{12} U_2 + U_1^2 + a_{11}^{12} U_1 U_2 + \delta_{12}^2 U_2^2
 + a_1^2 U_1^2 U_2 + a_1^1 \delta_{12} U_1 U_2^2 + U_1^2 U_2^2 ,
\nonumber \\
{\cal N}_1  & = &   1 + c_1^1 U_1 + c_1^2 \delta_{12} U_2 +  T_1 U_1^2 + c_{11}^{12} U_1 U_2 +  T_2 \delta_{12}^2 U_2^2
 + c_1^2 T_1 U_1^2 U_2 
\nonumber \\
& &   + c_1^1 T_2 \delta_{12} U_1 U_2^2+  T_1 T_2 U_1^2 U_2^2 ,
\nonumber \\
{\cal N}_2  & = &   1 + d_1^1 U_1 + d_1^2 \delta_{12} U_2 +  T_1 U_1^2 + d_{11}^{12} U_1 U_2 +  T_2 \delta_{12}^2 U_2^2
 + d_1^2 T_1 U_1^2 U_2
\nonumber \\
& &   + d_1^1 T_2 \delta_{12} U_1 U_2^2+  T_1 T_2 U_1^2 U_2^2 .
\label{4.19}
\end{eqnarray}
These results are fully consistent with Ref.~\cite{L7}. For $N=3$ (27 terms) the algorithm yields
\begin{eqnarray}
{\cal D} & = & 1 + b_1^1 U_1 + b_1^2 \delta_{12} U_2 + b_1^3 \delta_{13} \delta_{23} U_3 + U_1^2 + \delta_{12}^2 U_2^2+ \delta_{13}^2 
\delta_{23}^2 U_3^2+ b_{11}^{12} U_1 U_2
\nonumber \\ 
& &  + b_{11}^{13} \delta_{23} U_1 U_3 + b_{11}^{23} \delta_{12} \delta_{13} U_2 U_3+ b_1^2 U_1^2 U_2+ b_1^3 \delta_{23} U_1^2 U_3 + b_1^1 
\delta_{12} U_1 U_2^2
\nonumber \\
& & + b_1^1 \delta_{13}\delta_{23}^2 U_1 U_3^2 + b_1^2 \delta_{12} \delta_{13}^2 \delta_{23} U_2 U_3^2 + b_1^3 \delta_{12}^2 \delta_{13} 
U_2^2 U_3+ b_{111}^{123} U_1 U_2 U_3
\nonumber \\
& & + U_1^2 U_2^2+ \delta_{23}^2 U_1^2 U_3^2+ \delta_{12}^2 \delta_{13}^2 U_2^2 U_3^2+b_{11}^{23} U_1^2 U_2 U_3 + b_{11}^{12} \delta_{13} 
\delta_{23} U_1 U_2 U_3^2 
\nonumber \\
& & + 
b_{11}^{13} \delta_{12} U_1 U_2^2 U_3+ b_1^3 U_1^2 U_2^2 U_3 + b_1^2  \delta_{23} U_1^2 U_2 U_3^2 + b_1^1 \delta_{12} \delta_{13} U_1
 U_2^2 U_3^2 + U_1^2U_2^2U_3^2 ,
\nonumber \\
{\cal N} & = & 1 + a_1^1 U_1 + a_1^2 \delta_{12} U_2 + a_1^3 \delta_{13} \delta_{23} U_3 + U_1^2 + \delta_{12}^2 U_2^2+ \delta_{13}^2 
\delta_{23}^2 U_3^2+ a_{11}^{12} U_1 U_2
\nonumber \\ 
& &  + a_{11}^{13} \delta_{23} U_1 U_3 + a_{11}^{23} \delta_{12} \delta_{13} U_2 U_3+ a_1^2 U_1^2 U_2+ a_1^3 \delta_{23} U_1^2 U_3 + a_1^1 
\delta_{12} U_1 U_2^2
\nonumber \\
& & + a_1^1 \delta_{13}\delta_{23}^2 U_1 U_3^2 + a_1^2 \delta_{12} \delta_{13}^2 \delta_{23} U_2 U_3^2 + a_1^3 \delta_{12}^2 \delta_{13} 
U_2^2 U_3+ a_{111}^{123} U_1 U_2 U_3
\nonumber \\
& & + U_1^2 U_2^2+ \delta_{23}^2 U_1^2 U_3^2+ \delta_{12}^2 \delta_{13}^2 U_2^2 U_3^2+a_{11}^{23} U_1^2 U_2 U_3 + a_{11}^{12} \delta_{13} 
\delta_{23} U_1 U_2 U_3^2 
\nonumber \\
& & + 
a_{11}^{13} \delta_{12} U_1 U_2^2 U_3+ a_1^3 U_1^2 U_2^2 U_3 + a_1^2  \delta_{23} U_1^2 U_2 U_3^2 + a_1^1 \delta_{12} \delta_{13} U_1
 U_2^2 U_3^2 + U_1^2U_2^2U_3^2 ,
\nonumber \\
{\cal N}_1 & = & 1 + c_1^1 U_1 + c_1^2 \delta_{12} U_2 + c_1^3 \delta_{13} \delta_{23} U_3 + T_1 U_1^2 + T_2 \delta_{12}^2 U_2^2+ T_3 
\delta_{13}^2 \delta_{23}^2 U_3^2+ c_{11}^{12} U_1 U_2
\nonumber \\ 
& &  + c_{11}^{13} \delta_{23} U_1 U_3 + c_{11}^{23} \delta_{12} \delta_{13} U_2 U_3+ c_1^2 T_1 U_1^2 U_2+ c_1^3 T_1 \delta_{23} U_1^2 U_3 
+ c_1^1 T_2 \delta_{12} U_1 U_2^2
\nonumber \\
& & + c_1^1 T_3 \delta_{13}\delta_{23}^2 U_1 U_3^2 + c_1^2 T_3 \delta_{12} \delta_{13}^2 \delta_{23} U_2 U_3^2 + c_1^3 T_2 \delta_{12}^2
 \delta_{13} U_2^2 U_3+ c_{111}^{123} U_1 U_2 U_3
\nonumber \\
& & + T_1 T_2 U_1^2 U_2^2+ T_1 T_3 \delta_{23}^2 U_1^2 U_3^2+ T_2 T_3 \delta_{12}^2 \delta_{13}^2 U_2^2 U_3^2+ c_{11}^{23} T_1 U_1^2 U_2 U_3 
\nonumber \\
& & + c_{11}^{12} T_3 \delta_{13} \delta_{23} U_1 U_2 U_3^2 + c_{11}^{13} T_2 \delta_{12} U_1 U_2^2 U_3+ c_1^3 T_1 T_2 U_1^2 U_2^2 U_3 
\nonumber \\
& &  + c_1^2 T_1 T_3  \delta_{23} U_1^2 U_2 U_3^2 + c_1^1 T_2 T_3 \delta_{12} \delta_{13}  U_1 U_2^2 U_3^2 + T_1 T_2 T_3U_1^2U_2^2U_3^2 ,
\nonumber
\end{eqnarray}
\begin{eqnarray}
{\cal N}_2 & = & 1 + d_1^1 U_1 + d_1^2 \delta_{12} U_2 + d_1^3 \delta_{13} \delta_{23} U_3 + T_1 U_1^2 + T_2 \delta_{12}^2 U_2^2+ T_3 
\delta_{13}^2 \delta_{23}^2 U_3^2+ d_{11}^{12} U_1 U_2
\nonumber \\ 
& &  + d_{11}^{13} \delta_{23} U_1 U_3 + d_{11}^{23} \delta_{12} \delta_{13} U_2 U_3+ d_1^2 T_1 U_1^2 U_2+ d_1^3 T_1 \delta_{23} U_1^2 U_3 
+ d_1^1 T_2 \delta_{12} U_1 U_2^2
\nonumber \\
& & + d_1^1 T_3 \delta_{13}\delta_{23}^2 U_1 U_3^2 + d_1^2 T_3 \delta_{12} \delta_{13}^2 \delta_{23} U_2 U_3^2 + d_1^3 T_2 \delta_{12}^2
 \delta_{13} U_2^2 U_3+ d_{111}^{123} U_1 U_2 U_3
\nonumber \\
& & + T_1 T_2 U_1^2 U_2^2+ T_1 T_3 \delta_{23}^2 U_1^2 U_3^2+ T_2 T_3 \delta_{12}^2 \delta_{13}^2 U_2^2 U_3^2+ d_{11}^{23} T_1 U_1^2 U_2 U_3 
\nonumber \\
& & + d_{11}^{12} T_3 \delta_{13} \delta_{23} U_1 U_2 U_3^2 + d_{11}^{13} T_2 \delta_{12} U_1 U_2^2 U_3+ d_1^3 T_1 T_2 U_1^2 U_2^2 U_3 
\nonumber \\
& &  + d_1^2 T_1 T_3  \delta_{23} U_1^2 U_2 U_3^2 + d_1^1 T_2 T_3 \delta_{12} \delta_{13}  U_1 U_2^2 U_3^2 + T_1 T_2 T_3U_1^2U_2^2U_3^2 .
\label{4.20}
\end{eqnarray}

Inspection of these examples shows the following advantages of presenting results in the form of an algorithm. First, the recursion relations
relate $N$-baryon coefficients to $(N-1)$-baryon coefficients, cf. Eqs.~(\ref{4.6},\ref{4.9}). The algorithm gives directly the iterated result where 
everything is expressed
in terms of irreducible coefficients for 1,2,...,$N$ baryons. Secondly, the number of terms in the explicit expressions increases like $3^N$, so 
that writing down the explicit
expressions like in (\ref{4.19},\ref{4.20}) becomes quickly prohibitive. The algorithm on the other hand has been stated concisely for arbitrary 
$N$. It can also easily be implemented
in MAPLE, so that it is never necessary to deal manually with lengthy expressions. 

As a result of this section, we have reduced $S$ and $\psi_{\zeta}$ to those coefficients $a,b,c,d$ whose subscripts contain 
only 1's and which refer to 1,2,...,$N$ baryons with all permutations of labels. These irreducible coefficients have to be determined algebraically from 
the Dirac equation (\ref{3.15}) and are the
subject of the following section.
\section{Irreducible coefficients}\label{sect5}
We denote those $N$-baryon coefficients of the polynomials ${\cal N}, {\cal D}, {\cal N}_1, {\cal N}_2$ which cannot be 
determined recursively from the $N-1$ baryon problem as irreducible. As explained above, there are only 4 such coefficients for given $N$, 
namely the 
coefficients of the monomials $U_1 U_2... U_N$ in each of the 4 polynomials, $a_{11...1}^{12...N},b_{11...1}^{12...N},c_{11...1}^{12...N},
d_{11...1}^{12...N}$.
They encode the dynamical information about the situation where all $N$ baryons overlap and have to be determined by means of the 
Dirac equation.
For reasons to be discussed later in more detail, this is a difficult task for computer algebra programs like MAPLE, once the baryon
 number gets too large.
We have therefore determined the irreducible coefficients for low baryon numbers analytically, analyzed their structure and extrapolated 
the formulas
to arbitrary $N$. In this section we present our conjectured results for the 4 irreducible coefficients and general $N$. In the next section,
we will describe in detail the extent to which these conjectured results have actually been checked so far.

Given the complexity of the coefficients, it is once again easier for us to communicate our results in the form of an 
algorithm, rather than
a closed expression. The algorithm is actually a very simple one. Let us define a combinatorial expression ${\cal C}_N$ through the following
 two steps:
\begin{enumerate}
\item Write down the product
\begin{equation}
{\cal C}_N = \prod_{i<j}^N (1+B_{ij}),
\label{5.1}
\end{equation}
where $B_{ij}$ is a $N \times N$ matrix, and expand it.
\item For each of the $2^{N(N-1)/2}$ terms in the sum and each index $i=1,...,N$, denote by $n_i$ the number of indices $i$ 
appearing in this term ($n_i\leq N-1$). Then, if $k_i= N-1-n_i$ is odd, multiply the term by
\begin{equation}
R_i .
\label{5.2}
\end{equation}
\end{enumerate}
By way of example, we write down the explicit result for $N=2$ (2 terms),
\begin{equation}
{\cal C}_2 = R_1R_2 + B_{12},
\label{5.3}
\end{equation}
and  $N=3$ (8 terms),
\begin{eqnarray}
{\cal C}_3 & = &  1 + R_1R_2 B_{12} +R_1R_3 B_{13}+R_2R_3  B_{23}
\nonumber \\
& & + R_1 R_2 B_{13}B_{23} + R_1 R_3 B_{12}B_{23} + R_2 R_3 B_{12}B_{13} + B_{12}B_{13}B_{23}.
\label{5.4}
\end{eqnarray}
After this preparation, the irreducible coefficients can be expressed in compact form as follows,
\begin{eqnarray}
a_{11...1}^{12...N} & = & \frac{\prod_{i=1}^N a_1^i}{d_N} {\cal C}_N (R_i=\rho_i, B_{jk}),
\nonumber \\
b_{11...1}^{12...N} & = & \frac{\prod_{i=1}^N b_1^i}{d_N}{\cal C}_N (R_i=0, B_{jk}),
\nonumber \\
c_{11...1}^{12...N} & = &  \frac{\prod_{i=1}^N c_1^i}{d_N} {\cal C}_N (R_i=\mu_i, B_{jk}),
\nonumber \\
d_{11...1}^{12...N} & = &   \frac{\prod_{i=1}^N d_1^i}{d_N} {\cal C}_N (R_i=\nu_i,  B_{jk}),
\label{5.5}
\end{eqnarray}
with
\begin{equation}
d_N = \prod_{i<j} d_{ij}.
\label{5.6}
\end{equation}
All what remains to be done is to define exactly the various symbols appearing in (\ref{5.5},\ref{5.6}). We divide them into two categories.
The first category comprises those symbols which can be deduced from the single DHN baryon problem \cite{L3},
\begin{eqnarray}
a_1^i & = & - \frac{2(Z_i^4+1)}{Z_i(Z_i^2+1)},
\nonumber \\
b_1^i & = & - \frac{4Z_i}{Z_i^2+1},
\nonumber \\
c_1^i & = &  \frac{2 [ Z_i^4+1-2 \zeta_i^2 Z_i^2]}{(Z_i^2+1)( \zeta_i-Z_i)(\zeta_i Z_i +1)},
\nonumber \\
d_1^i & = &  \frac{2 [2 Z_i^2 - \zeta_i^2(Z_i^4+1)]}{(Z_i^2+1)(\zeta_i-Z_i)(\zeta_i Z_i +1)}.
\label{5.7}
\end{eqnarray}
They enter in the prefactor of the combinatorial expression ${\cal C}_N$ in Eq.~(\ref{5.5}) and are the same
as in Eqs.~(\ref{2.2},\ref{2.3}), up to trivial normalization factors in $c_1^i$ and $d_1^i$.  
The 2nd category consists of symbols which can be deduced from the two-baryon problem if one applies these formulas to $N=2$ and
compares them with the results of Ref.~\cite{L7},
\begin{eqnarray}
d_{ij} & = &  - 2 \frac{(\zeta_i Z_i- \zeta_j Z_j)(\zeta_j Z_i-\zeta_i Z_j)(\zeta_i Z_i Z_j+\zeta_j)(\zeta_jZ_iZ_j+ \zeta_i)}
{\zeta_i^2 \zeta_j^2 (Z_i^4-1)(Z_j^4-1)},
\nonumber \\
B_{ij} & = & \frac{2(\zeta_i^4+\zeta_j^4)Z_i^2 Z_j^2 - \zeta_i^2 \zeta_j^2 (Z_i^4+1)(Z_j^4+1)}{\zeta_i^2 \zeta_j^2 (Z_i^4-1)(Z_j^4-1)},
\nonumber \\
\rho_i & = & \frac{Z_i^4-1}{Z_i^4+1},
\nonumber \\
\mu_i & = &  \frac{Z_i^4-1}{Z_i^4+1-2\zeta_i^2 Z_i^2 },
\nonumber \\
\nu_i & = &  \frac{(Z_i^4-1)\zeta_i^2}{2 Z_i^2-\zeta_i^2(Z_i^4+1)}.
\label{5.8}
\end{eqnarray}
We have used everywhere the spectral parameter $\zeta_i$ boosted into the rest frame of baryon $i$, introduced in Eq.~(\ref{3.14}).
Note however that $\zeta_i$ could be replaced by $\eta_i$ in $d_{ij}$ and $B_{ij}$, so that the $\zeta$-dependence of these quantities
is spurious.
By using the variable $Z_i$ rather than $y_i$ and $\zeta_i$ rather than $v_i$ and $k$, we have achieved that all the basic expressions
are rational functions of the 2$N$ arguments ($Z_i,\zeta_i$). The same holds true for $\delta_{ij}$, Eq.~(\ref{4.2}), and $T_k$, Eq.~(\ref{4.7a}).

A noteworthy property of this construction is the fact that the algorithm leading to ${\cal C}_N$ is based on a factorization
in terms of quantities $B_{ij}$ referring to 2 baryons $i,j$ only, see Eq.~(\ref{5.1}). This implies that the solution of the two-baryon 
scattering problem is sufficient to determine completely $N$ baryon scattering. This observation is behind the phrase ``evidence for
factorized scattering" in the title of this paper. It goes beyond the usual factorization of the fermion scattering matrix, which holds trivially
in our case (see Sec.~\ref{sect9}). It teaches us that even when all $N$ baryons overlap, there is nothing new going on as compared to
having two overlapping baryons only. In this sense, factorization does not only hold for the on-shell scattering matrix, but also off-shell.

\section{Status of checking the above formulas}\label{sect6}
In the preceding sections, we have provided rules for explicitly constructing the scalar potential $S$ and the continuum spinors 
$\psi_{\zeta}$ for the $N$-baryon 
TDHF problem in the GN model. Let us summarize where we stand. The main ingredients in $S$ and $\psi_{\zeta}$ are 4 polynomials 
in $N$ exponentials $U_i$, consisting of $3^N$ terms each.
The coefficients in these polynomials can all be expressed through a set of irreducible coefficients multiplying $U_1 U_2...U_n$ in
 the $n$ baryon problem,
time delay factors $\delta_{ij}$ and fermion transmission amplitudes $T_i$, using the algorithm of Sec.~\ref{sect4}. The irreducible 
coefficients in turn
can be constructed starting from 1- and 2-baryon input only, using the algorithm of Sec.~\ref{sect5}. 

Since the Dirac equation
reduces to a set of algebraic equations and all ingredients are known rational functions, one would not expect any
 particular difficulties in checking that the spinor satisfies the Dirac equation, using computer algebra programs like MAPLE. 
However, the complexity of the resulting
expressions increases rapidly with increasing baryon number, quickly exceeding the capabilities of MAPLE due to storage and 
computation time problems.
Thus, for $N=2$ and $N=3$, we could still check all $2\times 5^N$ algebraic equations analytically with MAPLE in a straightforward way.
For $N=4$ or larger, the maximum size of expressions which MAPLE can handle is exceeded and  we have only been able 
to check our results numerically, for random values of the input parameters $Z_i,\zeta_i$.
This test has been carried out successfully for $N=4,...,8$.  By increasing the number of digits, one
can find out whether the floating point result is exact or approximate. Since the number of operations increases faster than
exponentially with $N$, it is actually necessary to run MAPLE with very high accuracy for large $N$ values. Thus for example,
during a full $N=8$ calculation, 40 digits get lost, so that one has to start out with 50 digits precision to be sure that the Dirac
equation is solved exactly.

Clearly, there must be a way of proving our results in full generality. The complexity of the solution and the intricate way in 
which $N$ baryon scattering is related to the scattering
problem of fewer baryons have prevented us so far from finding such a proof. Therefore, strictly speaking, our result still has the
status of a conjecture. In the  meantime, we shall restrict all applications shown below to 
problems with low values of $N$ for which we have established the validity beyond any doubt. We are confident that the results 
hold for arbitrary $N$, but this has to await a complete mathematical proof.

Up to this point, we have only dealt with the Dirac equation for continuum spinors. This still leaves open other aspects of the
 full TDHF problem like bound states,
self-consistency, and fermion density. In some sense, all we have achieved so far is to find time-dependent, transparent potentials 
for the Dirac equation, which
look asymptotically like boosted static potentials. This solves in part another open problem which has been 
raised in the literature \cite{L12}, namely to
classify all time-dependent, transparent potentials of the 1+1 dimensional Dirac equation.
How general is our result in this respect? All static transparent potentials are well known (see the discussion in Sec.~\ref{sect2}).
We can now construct all time-dependent transparent potentials which asymptotically  consist of an arbitrary number of 
such static solutions, boosted to
arbitrary velocities. This cannot be the complete set of all transparent potentials though, as evidenced by the example of the breather 
which does not fit into 
this scheme. Evidently, there must be another set of solutions where boosted breathers appear as asymptotic states, in addition to boosted 
static bound states. We do not know yet whether our ansatz will be capable of describing this more general class of solutions. All we
have checked is that the single breather can indeed be
reproduced with our ansatz, provided we allow for complex valued $U_i$'s. Scattering problems involving breathers are 
interesting in their own right, but will be left for future studies. 

\section{Bound states}\label{sect7}
In the $N$ baryon problem, one expects $N$ positive and $N$ negative energy bound states. As discussed in Ref.~\cite{L7},
the bound state spinors can be obtained 
from the continuum spinors by analytic continuation in the spectral parameter $\zeta$. To this end we first re-introduce the $\zeta$
dependence of the coefficients (\ref{5.5}--\ref{5.8}) by using $\zeta_i=\eta_i \zeta$. 
Only the coefficients $c_1^i,d_1^i, T_i, \mu_i, \nu_i$ are $\zeta$-dependent.
For positive energy bound states for example, $c_1^i, d_1^i, T_i$ develop a single pole at $\zeta=Z_i/\eta_i$. The bound state spinor
associated with baryon $i$ can then be obtained from the residue of $\psi_{\zeta}$ at the pole,
\begin{equation}
\psi^{(i)} =  N^{(i)} \lim_{\zeta \to Z_i/\eta_i} (\zeta \eta_i- Z_i) \psi_{\zeta}.
\label{7.1}
\end{equation}
The result is a normalizable solution of the Dirac equation.
The normalization factor $N^{(i)}$ can readily be determined for times $t$ when the $i$-th baryon is isolated, with the result
\begin{equation}
N^{(i)}  =  \frac{1}{2Z_i} \sqrt{ \frac{(Z_i^2+1)(Z_i^2+\eta_i^2)}{\eta_i (Z_i^2-1)}}
 \prod_{j(<i)} \delta_{ji}^{-1/2}.
\label{7.2}
\end{equation}
For this value of $N^{(i)}$, the bound state spinor (\ref{7.1}) is normalized according to
\begin{equation}
\int dx \psi^{(i)\dagger} \psi^{(i)} = 1.
\label{7.3}
\end{equation}
This method has been checked analytically for $N=2$ in Ref.~\cite{L7} and numerically for $N=3$ by us.
\section{Self-consistency and fermion density}\label{sect8}
The situation in the $N$-baryon problem is the same as in the 2-baryon problem \cite{L7}. The scalar density for a continuum state can
 be decomposed as
\begin{equation}
\bar{\psi}_{\zeta}\psi_{\zeta} = \left( \bar{\psi}_{\zeta}\psi_{\zeta} \right)_1 + \left( \bar{\psi}_{\zeta}\psi_{\zeta} \right)_2,
\label{8.1}
\end{equation}
where
\begin{equation}
\left( \bar{\psi}_{\zeta}\psi_{\zeta} \right)_1 = - \frac{2\zeta}{\zeta^2+1} S 
\label{8.2} 
\end{equation}
is the perturbative piece which gives self-consistency by itself. The 2nd part is cancelled against the discrete state
contribution,
\begin{equation}
\int_0^{\infty} \frac{d\zeta}{2\pi} \frac{\zeta^2+1}{2 \zeta^2}  \left( \bar{\psi}_{\zeta}\psi_{\zeta} \right)_2 = - \frac{i}{2\pi} \sum_{i=1}^N
\left( \bar{\psi}\psi \right)^{(i)} \ln Z_i^4,
\label{8.3}
\end{equation}
if one makes use of the self-consistency conditions in the asymptotic in- and out-states.
We can deduce $ \left( \bar{\psi}_{\zeta}\psi_{\zeta} \right)_2$ by subtracting the expression (\ref{8.2}) from the full scalar density
and can then check Eq.~(\ref{8.3}) numerically, since we know the discrete state spinors and the integral is convergent.
This test has been performed analytically for $N=2$ in Ref.~\cite{L7} and numerically for $N=3$ in the present work.

Likewise, the fermion density can be dealt with in the same manner as for 1 or 2 baryons. The basic identity is
\begin{equation}
\int_0^{\infty} \frac{d\zeta}{2\pi} \frac{\zeta^2+1}{2 \zeta^2} \left( \psi_{\zeta}^{\dagger}\psi_{\zeta} -1 \right) = - \sum_{i=1}^N \left(\psi^{\dagger}
\psi \right)^{(i)},
\label{8.4}
\end{equation}
relating the continuum and bound state densities \cite{L7}. The integral is convergent owing to the vacuum subtraction. 
We have checked this identity here numerically for $N=3$. From this and the self-consistency relation, one can again express the total, subtracted
fermion density through the bound state densities as
\begin{equation}
\rho = \sum_{i=1}^N \left(\nu_{i,+}-\nu_{i,-}-1 \right) \rho^{(i)},
\label{8.5} 
\end{equation}
generalizing the $N=2$ results \cite{L7}.
\section{Phase shifts, time delays and moduli}\label{sect9}
The fermion transmission amplitude for the $N$-baryon problem factorizes, since it can be evaluated when all baryons are far apart,
\begin{equation}
T = T_1 T_2 ... T_N,
\label{9.1}
\end{equation}
with $T_k$ from Eq.~(\ref{4.7a}). This fact has actually already been used in the normalization conditions (\ref{4.8}). The more 
interesting question is
how to characterize the outcome of the scattering process in terms of the baryon or multi-baryon bound states. 
Comparing the asymptotics for $t\to \pm \infty$, we find that the exponential $U_i$ acquires the following factor during an arbitrary  
$N$-baryon collision,
\begin{equation}
U_i \to U_i'= \left(\prod_{j (v_j<v_i)} \frac{1}{\delta_{ij}}\right) U_i  \left(\prod_{k (v_k>v_i)} \delta_{ki}\right) .
\label{9.2}
\end{equation}
The $\delta_{ij}$ have been given in Eq.~(\ref{4.2}).
If $v_j=v_i$ for one or several $j$'s, there is no shift factor because baryons $i$ and $j$ belong to the same compound state (``nucleus")
and do not scatter from each other. 

How does this translate into observables? The scattering process at the level of the TDHF potential
is classical, so that the situation is analogous to classical soliton scattering. If a single baryon is involved in the scattering process, the 
situation is very simple. The incoming and outgoing baryons can be associated with straight-line space-time trajectories defined by
\begin{equation}
\ln U_i = 0, \qquad \ln U_i' = 0.
\label{9.3}
\end{equation}
They have the same slope in the ($x,t$) diagram, since the velocity does not change. The factor $U_i'/U_i$ given in Eq.~(\ref{9.2}) then leads to 
a parallel shift of the outgoing space-time trajectory, which is usually interpreted as time delay (or advance). 

If an $n$-baryon bound state (``nucleus")
is scattered, the initial state contains $n$ baryon constituents $U_{i_1},...,U_{i_n}$ moving with the same velocity $v$ on parallel 
straight-line trajectories. Such a bound state depends on the scale factors $\lambda_i$ of $U_i$ (``moduli"), cf. Eq.~(\ref{3.6}), 
determining the
relative positions and the shape of the bound state without affecting its energy. In the final state, the $n$ trajectories will be displaced
laterally relative to the incoming trajectories. Since all $y$ parameters within one composite state must be chosen differently, according 
to (\ref{9.2}), the
displacement will be different for each trajectory. Therefore the net result cannot be interpreted anymore as a mere time delay, but is always 
accompanied by a change in moduli space, resulting in different relative baryon positions and a corresponding deformation of the scalar
potential. In this
sense, the scattering process is not really elastic and the composite bound states undergo a change in their internal structure. A time delay
of the full composite object could be defined, but this is neither unambiguous, nor necessary. The full asymptotic information about the
scattering process is contained in Eq.~(\ref{9.2}).
\section{Illustrative examples}\label{sect10}
Since we have verified the above formulas analytically or numerically with high precision for up to 8 baryons, we now present some
illustrative results for smaller values of $N$. Depending on the choice of 
velocity parameters, the same formalism can describe a variety of physical problems. 

For $N=2$, there are two distinct possibilities. If the  velocities
are chosen to be equal, we obtain a boosted 2-baryon bound state, provided that the $y$ parameters are different.
If the velocities are different, there is no restriction on the $y$ parameters and we describe scattering of baryon ($y_1,v_1$) on 
baryon ($y_2,v_2$).
In both cases, this yields nothing new as compared to Refs.~\cite{L3,L7}, but has been used to test our formulas. 

\begin{figure}[h]
\begin{center}
\epsfig{file=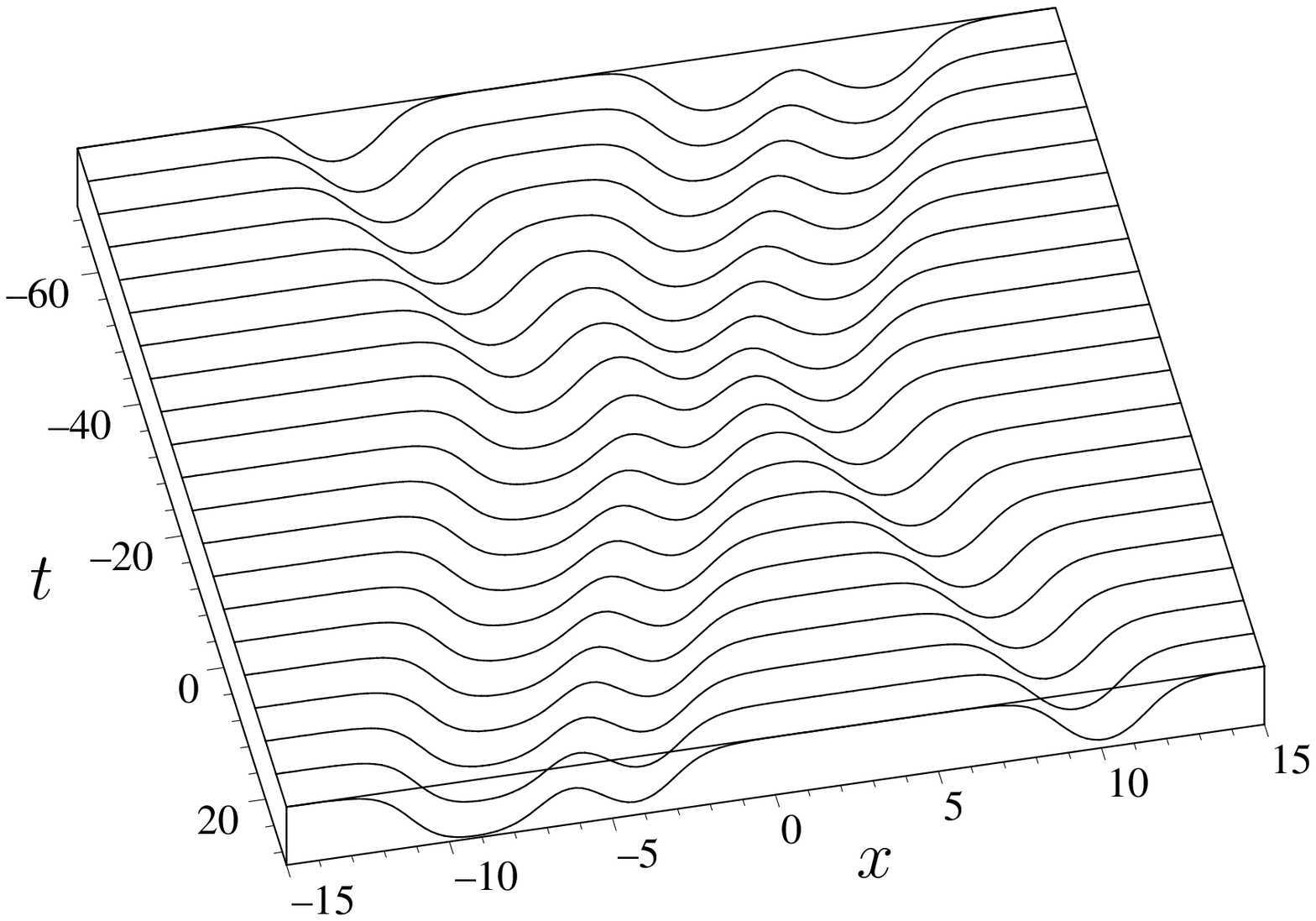,angle=270,width=8cm}
\caption{Example of baryon scattering from a 2-baryon bound state. The time evolution of the scalar TDHF potential is shown. Parameters:
$v_1=0.1,y_1=0.99, \lambda_1=1$ for the baryon, $v_2=v_3=-0.1, y_2=0.9999, y_3=0.9, \lambda_2=22.6, \lambda_3=0.06$ for the bound state.}
\label{fig7}
\end{center}
\end{figure}

\begin{figure}[h]
\begin{center}
\epsfig{file=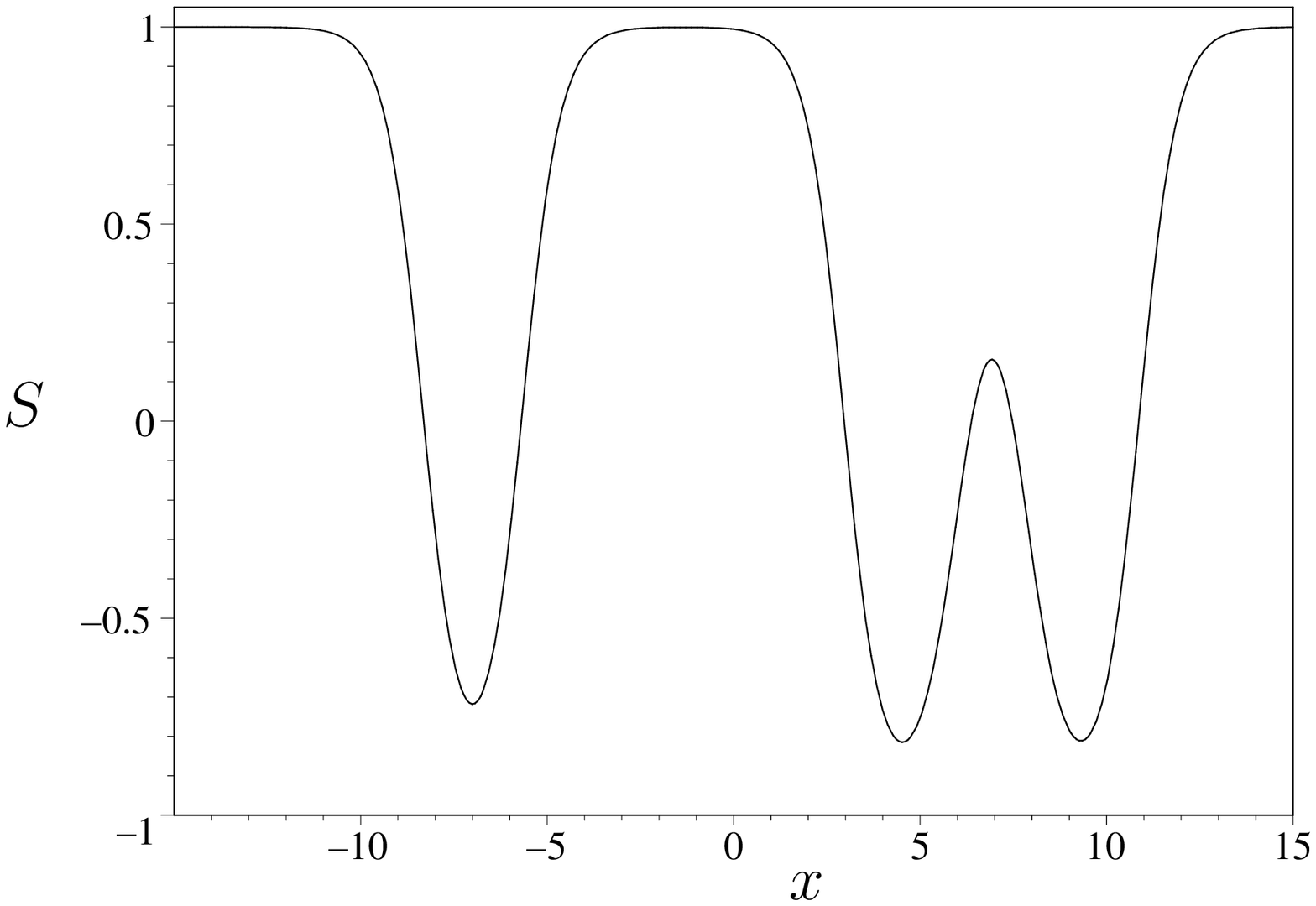,angle=270,width=6cm}\hskip 1.0cm \epsfig{file=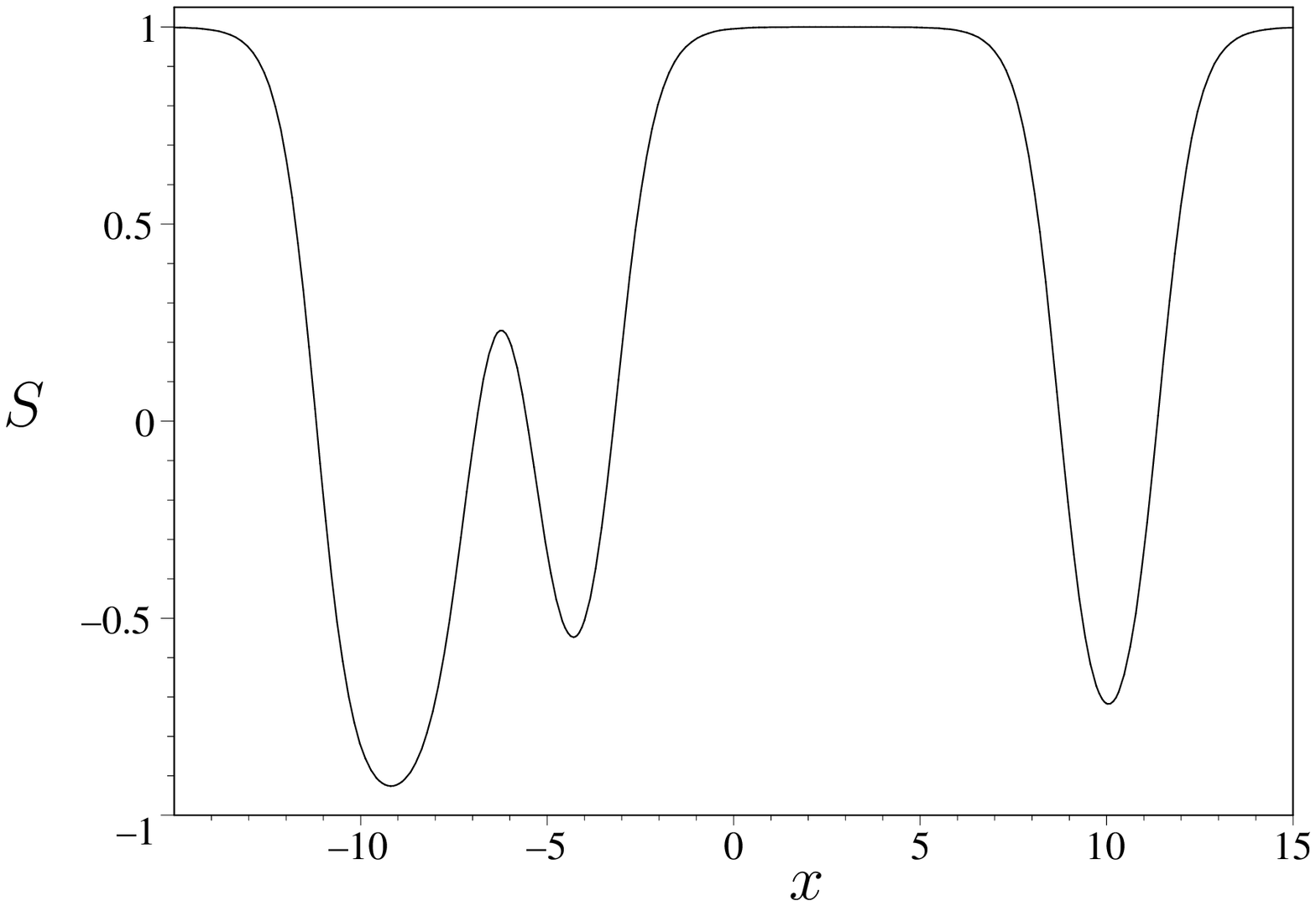,angle=270,width=6cm}
\caption{First and last frame of Fig. 7, showing the deformation of the 2-baryon bound state during the collision.}
\label{fig8}
\end{center}
\end{figure}

For $N=3$, we have to distinguish 
3 cases. If $v_1=v_2=v_3$ and all $y_i$'s are different, we are dealing with a boosted 3-baryon bound state. If two 
velocities are equal
and the corresponding $y$-parameters are different, the formalism describes scattering of a baryon on a 2-baryon bound state, 
analogous to 
$pd$-scattering in nature. An example of this process is shown in Fig.~\ref{fig7}, where the time evolution of the scalar TDHF potential 
during the collision is displayed. As announced above, the internal structure of the bound state necessarily changes during such a collision. 
To emphasize this point,
we compare in Fig.~\ref{fig8} the first and last time slice of Fig.~\ref{fig7}, i.e., the incoming and outgoing states. 
If all 3 velocities are different, the formalism describes a 3-baryon scattering process with 3 baryons in the initial and final state.
Since scattering processes with more than 2 incident particles are somewhat academic from the particle physics point of view, we do not show any 
example.

\begin{figure}[h]
\begin{center}
\epsfig{file=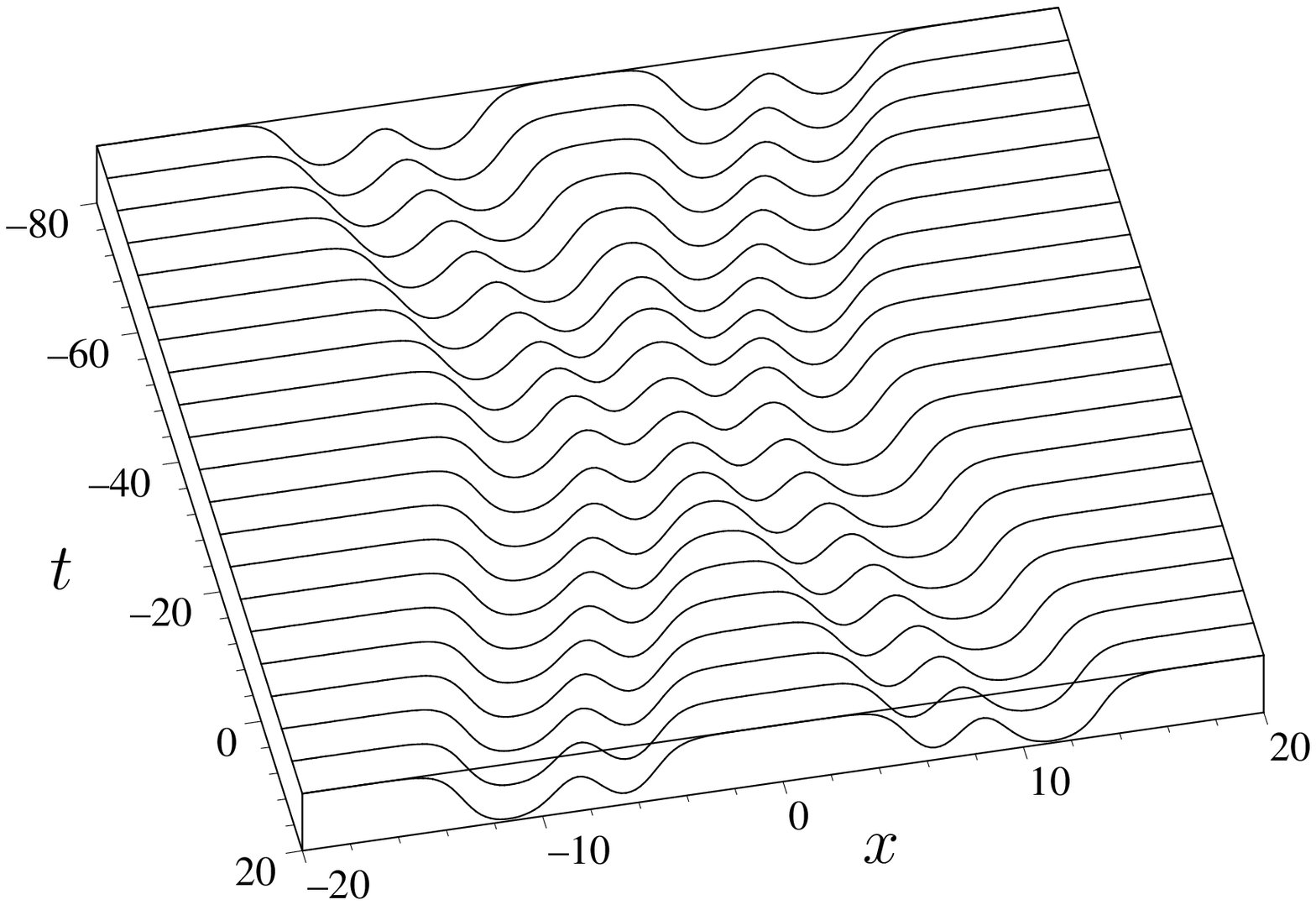,angle=270,width=8cm}
\caption{Example of scattering of two identical 2-baryon bound states, illustrated through the time evolution of the scalar TDHF potential $S$.
Parameters: Velocities $\pm 0.1$, bound state parameters: $y_1=0.9999, y_2=0.9, \lambda_1 =21.1 , \lambda_2 =0.064$.} 
\label{fig9}
\end{center}
\end{figure}

\begin{figure}[h]
\begin{center}
\epsfig{file=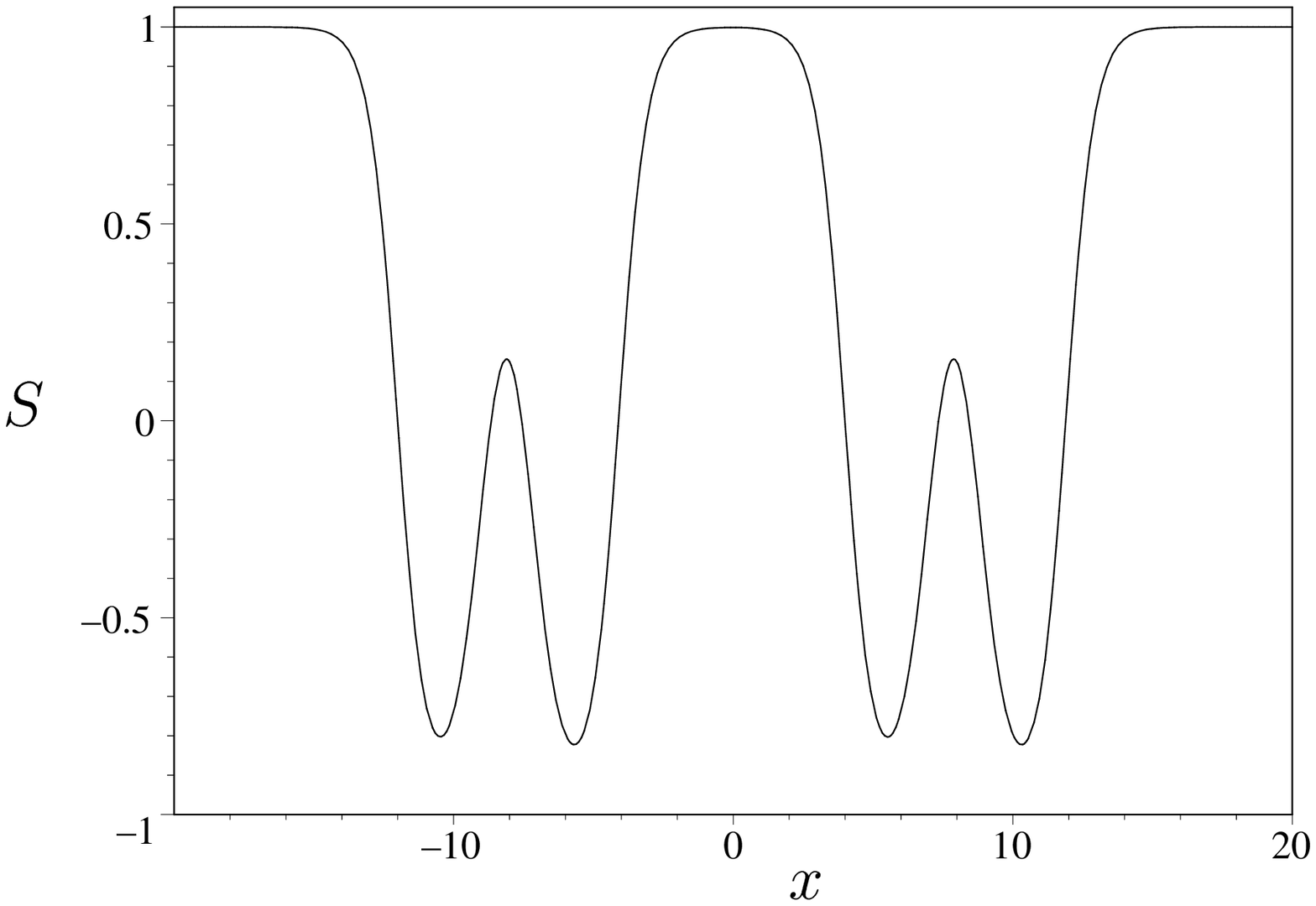,angle=270,width=6cm}\hskip 1.0cm \epsfig{file=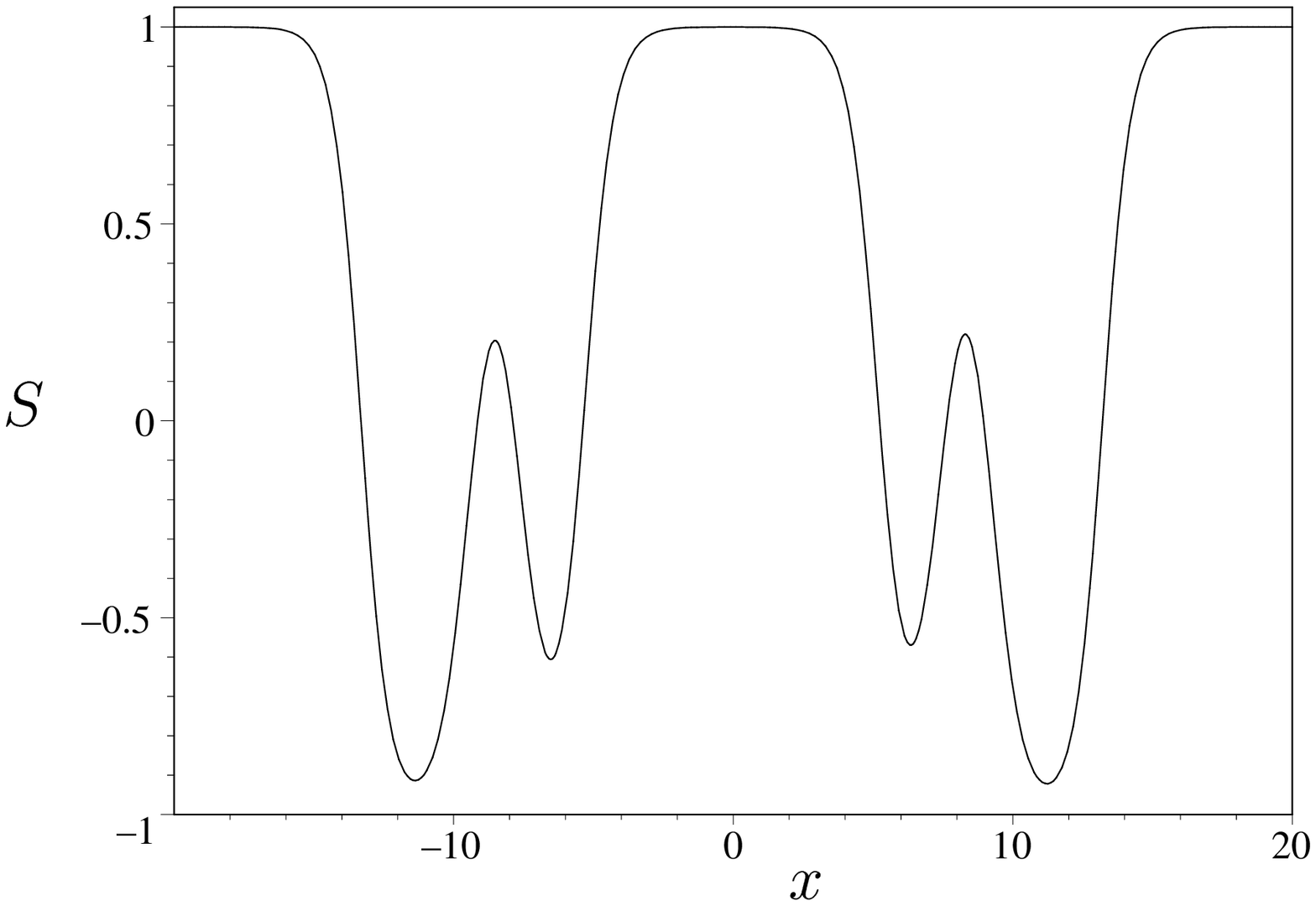,angle=270,width=6cm}
\caption{First and last frame of Fig. 9, to exhibit deformation of 2-baryon bound states as a result of the collision.}
\label{fig10}
\end{center}
\end{figure}

With increasing $N$, the number of scattering channels increases.
The next number of baryons is $N=4$, describing one boosted 4-baryon bound state, scattering of a baryon on a 3-baryon bound state, 
scattering of two 2-baryon bound
states, scattering of 3 particles (2 baryons and a 2-baryon bound state) or of 4 particles (4 individual baryons). The most interesting and new 
process out of these is
the scattering of 2 bound states, the analogue of $dd$-scattering --- the simplest case of nucleus-nucleus scattering. This is 
illustrated in Fig.~\ref{fig9}.
The change in structure of the bound state is exhibited more clearly in Fig.~\ref{fig10}.

\begin{figure}[h]
\begin{center}
\epsfig{file=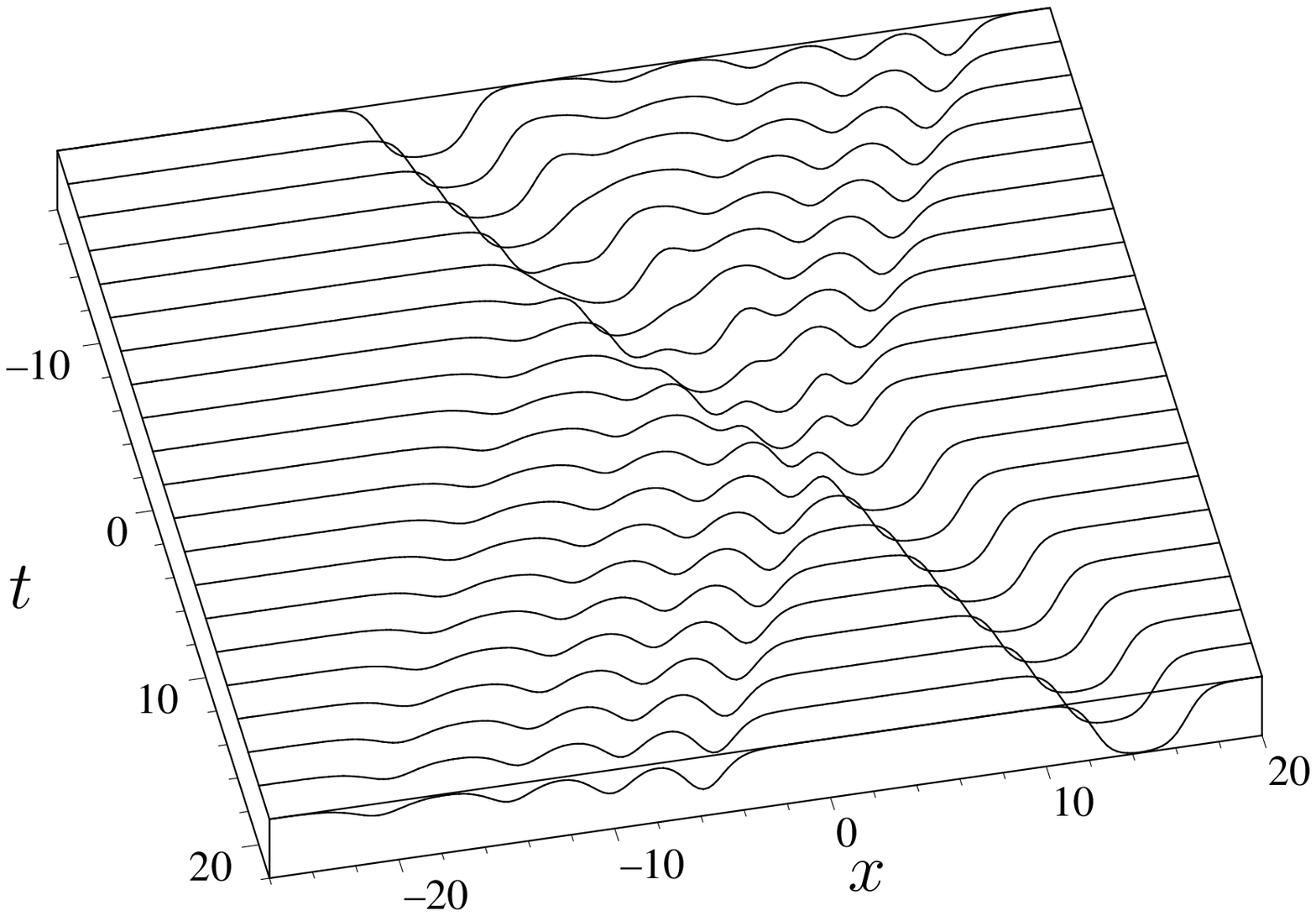,angle=270,width=8cm}
\caption{TDHF potential for scattering of a baryon ($y=0.9999,v=0.5$) on a 4-baryon bound state ($v=-0.5$). The bound state parameters are 
$y_1=0.9, y_2=0.8, y_3 = 0.7, y_4=0.6$ and all $\lambda_i=1$. Deformation of the bound state is less pronounced than in Figs.~\ref{fig7},\ref{fig9}
due to higher velocity.} 
\label{fig11}
\end{center}
\end{figure}

Finally, we give an example with 5 baryons. Out of the many possibilities, we have chosen scattering of a single baryon on a 4-baryon 
bound state,
the analogue of $p\alpha$-scattering in the real world, see Fig.~\ref{fig11}. We refrain from showing any results with larger number of 
baryons, since we have not yet checked our formulas thoroughly beyond $N=5$. However, we have no doubt that we could describe
correctly scattering processes with any number of baryons.

All of these examples involve topologically trivial bound states only. There is no difficulty in applying the same formulas to topologically 
non-trivial
scatterers as well. As already demonstrated in Ref.~\cite{L7}, all one has to do is let one or several $y$'s go to 1. Then, the corresponding 
baryon
becomes a  kink-antikink pair at infinite separation. This diverging separation has to be compensated by a change of the scale 
parameter $\lambda_i$
in the $U_i$ factor, so that half of the baryon disappears at infinity. In this way one can describe scattering of any number of topologically
trivial or non-trivial bound states, without need to derive separate formulas for this purpose.

\section{Summary and conclusions}\label{sect11}
This paper has dealt with the large $N$ limit of the GN model, the quantum field theory of massless, self-interacting, flavored fermions 
in 1+1 dimensions.
The fascinating aspect of Lagrangian (\ref{1.1}) is the fact that a single contact interaction term is able to generate a host of non-trivial 
phenomena. Even more surprisingly,
it seems that all of these can be worked out in closed analytical form, a rather exceptional situation in quantum field theory. 
The story begins with asymptotic freedom, the generation of a dynamical fermion mass, accompanied by spontaneous breakdown of 
the $Z_2$ chiral symmetry, 
and a scalar fermion-antifermion bound state, in the original work \cite{L1}. Soon afterwards baryons were 
discovered \cite{L5}, subsequently
complemented by a whole zoo of multi-baryon bound states \cite{L3}. As time evolved and computer algebra software became more 
powerful, ambitions
were raised, leading to results like soliton crystals in the ground state and phase diagram of dense matter \cite{L4} or time-dependent 
scattering processes of kinks and antikinks \cite{L16}. The most recent result is the TDHF solution of time-dependent baryon-baryon 
scattering \cite{L7}.

In the present work, we have tried to add another chapter to this progress report. By generalizing the joint ansatz for the TDHF potential 
and the spinors recently
proposed in Ref.~\cite{L7}, we have most probably found the solution to a whole class of scattering problems, namely all those where the 
incoming and
outgoing scatterers are boosted, static multi-fermion bound states of the GN model. The word ``probably" has to be used here because we 
have not yet been
able to prove our results in full generality. The solution which we have presented is based on the analytical solution of 
the 2- and 3-baryon problems,
followed by a tentative extrapolation to arbitrary $N$. These results have then been checked numerically for $N=4,...,8$, and all heralds
well for their general validity.
This method could only work because of a kind of factorization property which we have observed --- scattering of any number 
of baryons can 
apparently be predicted on the basis of 1- and 2-baryon input only. This holds not only for the asymptotic scattering data, but also 
during the
entire time evolution, where more than 2 baryons can overlap at a time. We interpret these findings as a large-$N$ manifestation of the 
quantum integrability of the GN  model.

The solution which we have presented is relevant for yet another problem, namely how to find transparent, time-dependent scalar
potentials for 
the Dirac equation in 1+1 dimensions.  It is clear that unlike in the static case, we have not yet arrived at the most general 
time-dependent solution. At least one time-dependent solution of the GN model is already known which does not belong 
to our class of solutions, the breather. It also yields a reflectionless potential. This suggests that a whole class of solutions is still
missing, namely the TDHF potentials of scattering processes involving breathers in the initial and final states. 
We know already that the single breather can be obtained with our ansatz if one admits complex valued exponentials
$U_i$. It will be interesting to see whether breather-baryon or breather-breather scattering can be solved along similar lines.

One other question which we have not been able to answer yet is whether our new solution is related to the solution of some known, 
classical non-linear equation
or system of equations. This question is a natural one, given prior experience. Thus for instance, all static baryons can be related to 
soliton solutions of the static NLS equation.
Higher bound states are related to the static multi-channel NLS equation. All dynamical kink solutions can be mapped onto
multi-soliton solutions 
of the sinh-Gordon equation. 
The non-relativistic limit of baryon-baryon scattering was shown to be equivalent to solutions of the time-dependent, 
multi-component NLS equation.
The advantage of such mappings is obvious. A lot of expertise and powerful techniques have been accumulated in the field of
non-linear systems over the years, which can be helpful for finding new solutions of the GN model or proving certain results in full generality.
A natural candidate for the present case would be the multi-component non-linear Dirac equation, i.e., the set of classical equations
\begin{equation}
\left( i \partial \!\!\!/ - \lambda  \sum_{k=1}^{n}\bar{\psi}_{k}\psi_{k}  \right)  \psi_{i} = 0.
\label{11.1}
\end{equation}
Inspection of the various condensates in Sec.~\ref{sect8} shows that it is indeed possible to construct solutions of Eq.~(\ref{11.1}) using 
our results. 
One needs $N+1$ components for $N$ baryons, since the solution is of type $N+1$. However, it is not possible to restrict oneself
to normalizable states as in the non-relativistic limit of the multi-component NLS equation. One would have to invoke $N$ different
bound states and one continuum state. Hence, even if our results are related to the classical system (\ref{11.1}), it seems very unlikely that
the solution presented here has already been given in the literature. Keeping a continuum state as one of the components would be very
hard to interpret classically. This is obviously a remnant of the Dirac sea, without analogue in the classical fermion system.

\section*{Acknowledgement}

We thank Gerald Dunne and Oliver Schnetz for stimulating discussions. This work has been supported in part by
the DFG under grant TH 842/1-1.

\end{document}